\input harvmac

\noblackbox

\let\includefigures=\iftrue
\let\useblackboard=\iftrue
\newfam\black

\includefigures
\message{If you do not have epsf.tex (to include figures),}
\message{change the option at the top of the tex file.}
\input epsf
\def\figin{\epsfcheck\figin}\def\figins{\epsfcheck\figins}
\def\epsfcheck{\ifx\epsfbox\UnDeFiNeD
\message{(NO epsf.tex, FIGURES WILL BE IGNORED)}
\gdef\figin##1{\vskip2in}\gdef\figins##1{\hskip.5in}
\else\message{(FIGURES WILL BE INCLUDED)}%
\gdef\figin##1{##1}\gdef\figins##1{##1}\fi}
\def\DefWarn#1{}
\def\figinsert{\goodbreak\midinsert}
\def\ifig#1#2#3{\DefWarn#1\xdef#1{Fig.~\the\figno}
\writedef{#1\leftbracket Fig.\noexpand~\the\figno}%
\figinsert\figin{\centerline{#3}}\medskip\centerline{\vbox{
\baselineskip12pt\advance\hsize by -1truein
\noindent\footnotefont{\bf Fig.~\the\figno:} #2}}
\bigskip\endinsert\global\advance\figno by1}
\else
\def\ifig#1#2#3{\xdef#1{Fig.~\the\figno}
\writedef{#1\leftbracket Fig.\noexpand~\the\figno}%
\global\advance\figno by1}
\fi

\def\doublefig#1#2#3#4{\DefWarn#1\xdef#1{Fig.~\the\figno}
\writedef{#1\leftbracket Fig.\noexpand~\the\figno}%
\figinsert\figin{\centerline{#3\hskip1.0cm#4}}\medskip\centerline{\vbox{
\baselineskip12pt\advance\hsize by -1truein
\noindent\footnotefont{\bf Fig.~\the\figno:} #2}}
\bigskip\endinsert\global\advance\figno by1}

\useblackboard
\message{If you do not have msbm (blackboard bold) fonts,}
\message{change the option at the top of the tex file.}
\font\blackboard=msbm10 scaled \magstep1
\font\blackboards=msbm7
\font\blackboardss=msbm5
\textfont\black=\blackboard
\scriptfont\black=\blackboards
\scriptscriptfont\black=\blackboardss

\else

\fi
%
\def\subsubsec#1{\bigskip\noindent{\it{#1}} \bigskip}
\def\yboxit#1#2{\vbox{\hrule height #1 \hbox{\vrule width #1
\vbox{#2}\vrule width #1 }\hrule height #1 }}
\def\fillbox#1{\hbox to #1{\vbox to #1{\vfil}\hfil}}
\def\ybox{{\lower 1.3pt \yboxit{0.4pt}{\fillbox{8pt}}\hskip-0.2pt}}
%
%


\def\comments#1{}

\def\tr{{\rm tr\ }}

\def\ket#1{|#1\rangle}

\def\vev#1{\langle{#1}\rangle}

\def\CT{{\cal T}}



\def\ap{\alpha'}

\def\II{\relax{I\kern-.10em I}}

\def\IZ{\relax{\rm Z\kern-.34em Z}}
\def\IB{\relax{\rm I\kern-.18em B}}
\def\IC{{\relax\hbox{$\inbar\kern-.3em{\rm C}$}}}
\def\ID{\relax{\rm I\kern-.18em D}}
\def\IE{\relax{\rm I\kern-.18em E}}
\def\IF{\relax{\rm I\kern-.18em F}}
\def\IG{\relax\hbox{$\inbar\kern-.3em{\rm G}$}}
\def\IGa{\relax\hbox{${\rm I}\kern-.18em\Gamma$}}
\def\IH{\relax{\rm I\kern-.18em H}}
\def\II{\relax{\rm I\kern-.18em I}}
\def\IK{\relax{\rm I\kern-.18em K}}
\def\IP{\relax{\rm I\kern-.18em P}}

%

\def\inbar{\,\vrule height1.5ex width.4pt depth0pt}

\def\IR{\relax{\rm I\kern-.18em R}}

\def\simgt{\hskip0.05in\relax{
\raise3.0pt\hbox{ $>$
{\lower5.0pt\hbox{\kern-1.05em $\sim$}} }} \hskip0.05in}

%


%

\def\lp10{\ell_p^{10}}
\def\lp11{\ell_p^{11}}
\def\R11{R_{11}}

\def\frac#1#2{{#1 \over #2}}



\newdimen\tableauside\tableauside=1.0ex
\newdimen\tableaurule\tableaurule=0.4pt
\newdimen\tableaustep
\def\phantomhrule#1{\hbox{\vbox to0pt{\hrule height\tableaurule width#1\vss}}}
\def\phantomvrule#1{\vbox{\hbox to0pt{\vrule width\tableaurule height#1\hss}}}
\def\sqr{\vbox{%
  \phantomhrule\tableaustep
  \hbox{\phantomvrule\tableaustep\kern\tableaustep\phantomvrule\tableaustep}%
  \hbox{\vbox{\phantomhrule\tableauside}\kern-\tableaurule}}}
\def\squares#1{\hbox{\count0=#1\noindent\loop\sqr
  \advance\count0 by-1 \ifnum\count0>0\repeat}}
\def\tableau#1{\vcenter{\offinterlineskip
  \tableaustep=\tableauside\advance\tableaustep by-\tableaurule
  \kern\normallineskip\hbox
    {\kern\normallineskip\vbox
      {\gettableau#1 0 }%
     \kern\normallineskip\kern\tableaurule}%
  \kern\normallineskip\kern\tableaurule}}
\def\gettableau#1 {\ifnum#1=0\let\next=\null\else
  \squares{#1}\let\next=\gettableau\fi\next}

\tableauside=1.0ex
\tableaurule=0.4pt


 %
 %
 \def\eqnn#1{\xdef #1{(\secsym\the\meqno)}\writedef{#1\leftbracket#1}%
 \global\advance\meqno by1\wrlabeL#1}
 \def\eqna#1{\xdef #1##1{\hbox{$(\secsym\the\meqno##1)$}}
 \writedef{#1\numbersign1\leftbracket#1{\numbersign1}}%
 \global\advance\meqno by1\wrlabeL{#1$\{\}$}}
 \def\eqn#1#2{\xdef #1{(\secsym\the\meqno)}\writedef{#1\leftbracket#1}%
 \global\advance\meqno by1$$#2\eqno#1\eqlabeL#1$$}

\global\newcount\itemno \global\itemno=0

\def\itemaut#1{\global\advance\itemno by1\noindent\item{\the\itemno.}#1}


\def\eg{{\it e.g.}}
\def\ie{{\it i.e.}}

\hyphenation{Di-men-sion-al}



\lref\Schomerusanomaly{
  S.~Fredenhagen and V.~Schomerus,
  ``On minisuperspace models of S-branes,''
  JHEP {\bf 0312}, 003 (2003)
  [arXiv:hep-th/0308205].
}

\lref\usnext{G. Horowitz, J. McGreevy, and E. Silverstein, in progress}

\lref\AtickWitten{
  J.~J.~Atick and E.~Witten,
``The Hagedorn Transition And The Number Of Degrees Of Freedom Of String
Theory,''
  Nucl.\ Phys.\ B {\bf 310}, 291 (1988).
}

\lref\BarbonDD{
  J.~L.~F.~Barbon and E.~Rabinovici,
  ``Touring the Hagedorn ridge,''
  arXiv:hep-th/0407236.
}

\lref\DeAlwisPR{
  S.~P.~de Alwis, J.~Polchinski and R.~Schimmrigk,
  ``Heterotic Strings With Tree Level Cosmological Constant,''
  Phys.\ Lett.\ B {\bf 218}, 449 (1989).
}

\lref\KachruED{
  S.~Kachru, J.~Kumar and E.~Silverstein,
  ``Orientifolds, RG flows, and closed string tachyons,''
  Class.\ Quant.\ Grav.\  {\bf 17}, 1139 (2000)
  [arXiv:hep-th/9907038].
}

\lref\HarveyNA{
  J.~A.~Harvey, D.~Kutasov and E.~J.~Martinec,
 ``On the relevance of tachyons,''
  arXiv:hep-th/0003101.
}

\lref\Matrixcosmo{
  J.~L.~Karczmarek and A.~Strominger,
``Matrix cosmology,''
  JHEP {\bf 0404}, 055 (2004)
  [arXiv:hep-th/0309138];
  J.~L.~Karczmarek and A.~Strominger,
``Closed string tachyon condensation at c = 1,''
  JHEP {\bf 0405}, 062 (2004)
  [arXiv:hep-th/0403169].
  J.~L.~Karczmarek, A.~Maloney and A.~Strominger,
``Hartle-Hawking vacuum for c = 1 tachyon condensation,''
  JHEP {\bf 0412}, 027 (2004)
  [arXiv:hep-th/0405092];
  S.~R.~Das and J.~L.~Karczmarek,
``Spacelike boundaries from the c = 1 matrix model,''
  Phys.\ Rev.\ D {\bf 71}, 086006 (2005)
  [arXiv:hep-th/0412093];
  S.~Hirano,
``Energy quantisation in bulk bouncing tachyon,''
  arXiv:hep-th/0502199.
 }

\lref\Gary{
  T.~Hertog and G.~T.~Horowitz,
``Towards a big crunch dual,''
  JHEP {\bf 0407}, 073 (2004)
  [arXiv:hep-th/0406134].
}

\lref\teschner{
  J.~Teschner,
 ``Liouville theory revisited,''
  Class.\ Quant.\ Grav.\  {\bf 18}, R153 (2001)
  [arXiv:hep-th/0104158].
}

\lref\BD{
N.D. Birrell, P.C.W. Davies,
{\it Quantum Fields in Curved Space},
 Cambridge, Uk: Univ.\ Pr.\ (1982) 340p.
}

\lref\Strompart{
  A.~Strominger,
 ``Open string creation by S-branes,''
  arXiv:hep-th/0209090;
  A.~Maloney, A.~Strominger and X.~Yin,
  ``S-brane thermodynamics,''
  JHEP {\bf 0310}, 048 (2003)
  [arXiv:hep-th/0302146].
}

\lref\StromGut{
  M.~Gutperle and A.~Strominger,
  ``Timelike boundary Liouville theory,''
  Phys.\ Rev.\ D {\bf 67}, 126002 (2003)
  [arXiv:hep-th/0301038].
}

\lref\StromTak{
  A.~Strominger and T.~Takayanagi,
``Correlators in timelike bulk Liouville theory,''
  Adv.\ Theor.\ Math.\ Phys.\  {\bf 7}, 369 (2003)
  [arXiv:hep-th/0303221].
}

\lref\SchomerusVV{
  V.~Schomerus,
  ``Rolling tachyons from Liouville theory,''
  JHEP {\bf 0311}, 043 (2003)
  [arXiv:hep-th/0306026].
}


\lref\LFTpartition{
  A.~Gupta, S.~P.~Trivedi and M.~B.~Wise,
``Random Surfaces In Conformal Gauge,''
  Nucl.\ Phys.\ B {\bf 340}, 475 (1990);
  M.~Bershadsky and I.~R.~Klebanov,
``Genus One Path Integral In Two-Dimensional Quantum Gravity,''
  Phys.\ Rev.\ Lett.\  {\bf 65}, 3088 (1990).
}

\lref\seiberg{
 N.~Seiberg,
``Notes On Quantum Liouville Theory And Quantum Gravity,''
  Prog.\ Theor.\ Phys.\ Suppl.\  {\bf 102}, 319 (1990).
}

\lref\GinspargIS{
  P.~H.~Ginsparg and G.~W.~Moore,
``Lectures on 2-D gravity and 2-D string theory,''
  arXiv:hep-th/9304011.
}

\lref\Poly{
  A.~M.~Polyakov,
  ``A few projects in string theory,''
  arXiv:hep-th/9304146.
}

\lref\APS{ A.~Adams, J.~Polchinski and E.~Silverstein, ``Don't panic! Closed string tachyons in ALE
space-times,'' JHEP {\bf 0110}, 029 (2001) [arXiv:hep-th/0108075].
}

\lref\TFA{
  A.~Adams, X.~Liu, J.~McGreevy, A.~Saltman and E.~Silverstein,
  ``Things fall apart: Topology change from winding tachyons,''
  arXiv:hep-th/0502021.
}

\lref\othertach{
  M.~Headrick, S.~Minwalla and T.~Takayanagi,
``Closed string tachyon condensation: An overview,''
  Class.\ Quant.\ Grav.\  {\bf 21}, S1539 (2004)
  [arXiv:hep-th/0405064];
  E.~J.~Martinec,
``Defects, decay, and dissipated states,''
  arXiv:hep-th/0210231.
}

\lref\garybubbles{G. Horowitz, "Tachyon Condensation and Black Strings," to appear}

\lref\CostaEJ{
  M.~S.~Costa, C.~A.~R.~Herdeiro, J.~Penedones and N.~Sousa,
  ``Hagedorn transition and chronology protection in string theory,''
  arXiv:hep-th/0504102;
A.~Adams and A.~Maloney, in progress.
}

\lref\stromcon{
A.~Strominger,
``Massless black holes and conifolds in string theory,'' Nucl.\ Phys.\ B {\bf 451}, 96 (1995)
[arXiv:hep-th/9504090].
}

\lref\dealwisetal{
S.~P.~de Alwis, J.~Polchinski and R.~Schimmrigk, ``Heterotic Strings With Tree Level Cosmological Constant,''
Phys.\ Lett.\ B {\bf 218}, 449 (1989);
R.~C.~Myers, ``New Dimensions For Old Strings,'' Phys.\ Lett.\ B {\bf 199}, 371 (1987).
}

\lref\DaCunhaFM{
J.~Polchinski,
``A Two-Dimensional Model For Quantum Gravity,''
Nucl.\ Phys.\ B {\bf 324}, 123 (1989);
B.~C.~Da Cunha and E.~J.~Martinec, ``Closed string tachyon condensation
and worldsheet inflation,'' Phys.\ Rev.\ D {\bf 68}, 063502 (2003) [arXiv:hep-th/0303087];
E.~J.~Martinec,
``The annular report on non-critical string theory,''
arXiv:hep-th/0305148.
}

\lref\WittenGJ{ E.~Witten, ``Instability Of The Kaluza-Klein Vacuum,'' Nucl.\ Phys.\ B {\bf 195}, 481 (1982).
}

\lref\ShankarCM{ R.~Shankar and E.~Witten, ``The S Matrix Of The Supersymmetric Nonlinear Sigma Model,'' Phys.\
Rev.\ D {\bf 17}, 2134 (1978).
}

\lref\AhnGN{ C.~Ahn, D.~Bernard and A.~LeClair, ``Fractional Supersymmetries In Perturbed Coset Cfts And
Integrable Soliton Theory,'' Nucl.\ Phys.\ B {\bf 346}, 409 (1990).
}

\lref\openconfine{
P.~Yi, ``Membranes from five-branes and fundamental strings from Dp branes,'' Nucl.\ Phys.\ B {\bf 550}, 214
(1999) [arXiv:hep-th/9901159].
O.~Bergman, K.~Hori and P.~Yi, ``Confinement on the brane,'' Nucl.\ Phys.\ B {\bf 580}, 289 (2000)
[arXiv:hep-th/0002223].
A.~Sen, ``Supersymmetric world-volume action for non-BPS D-branes,'' JHEP {\bf 9910}, 008 (1999)
[arXiv:hep-th/9909062].
}

\lref\sen{
A.~Sen, ``Non-BPS states and branes in string theory,'' arXiv:hep-th/9904207.
}

\lref\branedecay{
N.~Lambert, H.~Liu and J.~Maldacena, ``Closed strings from decaying D-branes,''
arXiv:hep-th/0303139;
  J.~L.~Karczmarek, H.~Liu, J.~Maldacena and A.~Strominger,
 ``UV finite brane decay,''
  JHEP {\bf 0311}, 042 (2003)
  [arXiv:hep-th/0306132].
}

\lref\otherRG{
J.~R.~David, M.~Gutperle, M.~Headrick and S.~Minwalla, ``Closed string tachyon condensation on twisted
circles,'' JHEP {\bf 0202}, 041 (2002) [arXiv:hep-th/0111212];
M.~Headrick, S.~Minwalla and T.~Takayanagi, ``Closed string tachyon condensation: An overview,'' Class.\ Quant.\
Grav.\  {\bf 21}, S1539 (2004) [arXiv:hep-th/0405064];
}

\lref\davetal{
D.~R.~Morrison and K.~Narayan, ``On tachyons, gauged linear sigma models, and flip transitions,''
arXiv:hep-th/0412337.
}

\lref\chicago{
 J.~A.~Harvey, D.~Kutasov, E.~J.~Martinec and G.~W.~Moore, ``Localized tachyons and RG flows,''
arXiv:hep-th/0111154;
}

\lref\FQS{
D.~Friedan, Z.~Qiu and S.~H.~Shenker, ``Conformal Invariance, Unitarity And Two-Dimensional Critical
Exponents,'' Phys.\ Rev.\ Lett.\  {\bf 52}, 1575 (1984).
}

\lref\KMS{
D.~A.~Kastor, E.~J.~Martinec and S.~H.~Shenker, ``Rg Flow In N=1 Discrete Series,'' Nucl.\ Phys.\ B {\bf 316},
590 (1989).
}

\lref\earlierSStach{
S.~Kachru, J.~Kumar and E.~Silverstein, ``Orientifolds, RG flows, and closed string tachyons,'' Class.\ Quant.\
Grav.\  {\bf 17}, 1139 (2000) [arXiv:hep-th/9907038].
}

\lref\ColemanBU{ S.~R.~Coleman, ``Quantum Sine-Gordon Equation As The Massive Thirring Model,'' Phys.\ Rev.\ D
{\bf 11}, 2088 (1975).
}

\lref\GreeneYB{ B.~R.~Greene, K.~Schalm and G.~Shiu, ``Dynamical topology change in M theory,'' J.\ Math.\
Phys.\  {\bf 42}, 3171 (2001) [arXiv:hep-th/0010207];
}

\lref\classtop{
P.~S.~Aspinwall, B.~R.~Greene and D.~R.~Morrison, ``Calabi-Yau moduli space, mirror manifolds and spacetime
topology  change in string theory,'' Nucl.\ Phys.\ B {\bf 416}, 414 (1994) [arXiv:hep-th/9309097];
E.~Witten, ``Phases of N = 2 theories in two dimensions,'' Nucl.\ Phys.\ B {\bf 403}, 159 (1993)
[arXiv:hep-th/9301042];
J.~Distler and S.~Kachru, ``(0,2) Landau-Ginzburg theory,'' Nucl.\ Phys.\ B {\bf 413}, 213 (1994)
[arXiv:hep-th/9309110].
J.~Distler and S.~Kachru, ``Duality of (0,2) string vacua,'' Nucl.\ Phys.\ B {\bf 442}, 64 (1995)
[arXiv:hep-th/9501111].
}

\lref\trapping{
L.~Kofman, A.~Linde, X.~Liu, A.~Maloney, L.~McAllister and E.~Silverstein, ``Beauty is attractive: Moduli
trapping at enhanced symmetry points,'' JHEP {\bf 0405}, 030 (2004) [arXiv:hep-th/0403001].
}

\lref\quantop{
B.~R.~Greene, D.~R.~Morrison and A.~Strominger, ``Black hole condensation and the unification of string vacua,''
Nucl.\ Phys.\ B {\bf 451}, 109 (1995) [arXiv:hep-th/9504145];
P.~Candelas and X.~C.~de la Ossa, ``Comments On Conifolds,'' Nucl.\ Phys.\ B {\bf 342}, 246 (1990);
S.~Kachru and E.~Silverstein, ``Chirality-changing phase transitions in 4d string vacua,'' Nucl.\ Phys.\ B {\bf
504}, 272 (1997) [arXiv:hep-th/9704185].
S.~Gukov, J.~Sparks and D.~Tong, ``Conifold transitions and five-brane condensation in M-theory on Spin(7)
Class.\ Quant.\ Grav.\  {\bf 20}, 665 (2003) [arXiv:hep-th/0207244].
}

\lref\MandelstamHB{
S.~Mandelstam,
``Soliton Operators For The Quantized Sine-Gordon Equation,''
Phys.\ Rev.\ D {\bf 11}, 3026 (1975).
}
\lref\ColemanBU{
S.~R.~Coleman,
``Quantum Sine-Gordon Equation As The Massive Thirring Model,''
Phys.\ Rev.\ D {\bf 11}, 2088 (1975).
}

\lref\KosterlitzXP{
J.~M.~Kosterlitz and D.~J.~Thouless,
``Ordering, Metastability And Phase Transitions In Two-Dimensional  Systems,''
J.\ Phys.\ C {\bf 6}, 1181 (1973).
}

\lref\KogutSN{
J.~B.~Kogut and L.~Susskind,
``Vacuum Polarization And The Absence Of Free Quarks In Four-Dimensions,''
Phys.\ Rev.\ D {\bf 9}, 3501 (1974).
}

\lref\PolyakovFU{
A.~M.~Polyakov,
``Quark Confinement And Topology Of Gauge Groups,''
Nucl.\ Phys.\ B {\bf 120}, 429 (1977).
}

\lref\AhnUQ{ C.~Ahn, ``Complete S Matrices Of Supersymmetric Sine-Gordon Theory And Perturbed Superconformal
Minimal Model,'' Nucl.\ Phys.\ B {\bf 354}, 57 (1991).
}

\lref\ZamolodchikovXM{ A.~B.~Zamolodchikov and A.~B.~Zamolodchikov, ``Factorized S-Matrices In Two Dimensions As
The Exact Solutions Of  Certain Relativistic Quantum Field Models,'' Annals Phys.\  {\bf 120}, 253 (1979).
}

\lref\PolchinskiFN{
J.~Polchinski,
``A Two-Dimensional Model For Quantum Gravity,''
Nucl.\ Phys.\ B {\bf 324}, 123 (1989).
}

\lref\Xiao{
S.~Hellerman and X.~Liu,
``Dynamical dimension change in supercritical string theory,''
arXiv:hep-th/0409071.
}

\lref\SaltmanJH{
A.~Saltman and E.~Silverstein,
``A new handle on de Sitter compactifications,''
arXiv:hep-th/0411271.
}

\lref\ScherkTA{
J.~Scherk and J.~H.~Schwarz,
``Spontaneous Breaking Of Supersymmetry Through Dimensional Reduction,''
Phys.\ Lett.\ B {\bf 82}, 60 (1979).
}

\lref\HollowoodEX{
T.~J.~Hollowood and E.~Mavrikis,
``The N = 1 supersymmetric bootstrap and Lie algebras,''
Nucl.\ Phys.\ B {\bf 484}, 631 (1997)
[arXiv:hep-th/9606116].
}

\lref\BajnokDK{
Z.~Bajnok, C.~Dunning, L.~Palla, G.~Takacs and F.~Wagner,
``SUSY sine-Gordon theory as a perturbed conformal field theory and  finite
size effects,''
Nucl.\ Phys.\ B {\bf 679}, 521 (2004)
[arXiv:hep-th/0309120].
}

\lref\FerraraJV{
S.~Ferrara, L.~Girardello and S.~Sciuto,
``An Infinite Set Of Conservation Laws Of The Supersymmetric Sine-Gordon
Theory,''
Phys.\ Lett.\ B {\bf 76}, 303 (1978).
}

\lref\WittenYC{
E.~Witten,
``Phases of N = 2 theories in two dimensions,''
Nucl.\ Phys.\ B {\bf 403}, 159 (1993)
[arXiv:hep-th/9301042].
}

\lref\VafaRA{
C.~Vafa,
``Mirror symmetry and closed string tachyon condensation,''
arXiv:hep-th/0111051.
}

\lref\MorrisonJA{
D.~R.~Morrison and K.~Narayan,
``On tachyons, gauged linear sigma models, and flip transitions,''
arXiv:hep-th/0412337.
}

\lref\BuscherQJ{
T.~H.~Buscher,
``Path Integral Derivation Of Quantum Duality In Nonlinear Sigma Models,''
Phys.\ Lett.\ B {\bf 201}, 466 (1988).
}

\lref\RocekPS{
M.~Rocek and E.~Verlinde,
``Duality, quotients, and currents,''
Nucl.\ Phys.\ B {\bf 373}, 630 (1992)
[arXiv:hep-th/9110053].
}

\lref\MorrisonYH{
D.~R.~Morrison and M.~R.~Plesser,
``Towards mirror symmetry as duality for two dimensional abelian gauge
theories,''
Nucl.\ Phys.\ Proc.\ Suppl.\  {\bf 46}, 177 (1996)
[arXiv:hep-th/9508107].
}

\lref\HoriKT{
K.~Hori and C.~Vafa,
``Mirror symmetry,''
arXiv:hep-th/0002222.
}

\lref\AharonySX{
O.~Aharony, J.~Marsano, S.~Minwalla, K.~Papadodimas and M.~Van Raamsdonk,
``The Hagedorn/deconfinement phase transition in weakly coupled large N gauge
theories,''
arXiv:hep-th/0310285.
}

\lref\ZamolodchikovGT{
A.~B.~Zamolodchikov,
``'Irreversibility' Of The Flux Of The Renormalization Group In A 2-D Field
Theory,''
JETP Lett.\  {\bf 43}, 730 (1986)
[Pisma Zh.\ Eksp.\ Teor.\ Fiz.\  {\bf 43}, 565 (1986)].
}

\lref\BanksQS{
T.~Banks and E.~J.~Martinec,
``The Renormalization Group And String Field Theory,''
Nucl.\ Phys.\ B {\bf 294}, 733 (1987).
}

\lref\DeAlwisKP{
  S.~P.~De Alwis and A.~T.~Flournoy,
``Closed string tachyons and semi-classical instabilities,''
  Phys.\ Rev.\ D {\bf 66}, 026005 (2002)
  [arXiv:hep-th/0201185].
}

\lref\LMS{
  H.~Liu, G.~W.~Moore and N.~Seiberg,
  ``Strings in a time-dependent orbifold,''
  JHEP {\bf 0206}, 045 (2002)
  [arXiv:hep-th/0204168];
and see references in
  L.~Cornalba and M.~S.~Costa,
``Time-dependent orbifolds and string cosmology,''
  Fortsch.\ Phys.\  {\bf 52}, 145 (2004)
  [arXiv:hep-th/0310099].
}
\lref\GaryJoeetal{
  G.~T.~Horowitz and J.~Polchinski,
  ``Instability of spacelike and null orbifold singularities,''
  Phys.\ Rev.\ D {\bf 66}, 103512 (2002)
  [arXiv:hep-th/0206228];
  A.~Lawrence,
  ``On the instability of 3D null singularities,''
  JHEP {\bf 0211}, 019 (2002)
  [arXiv:hep-th/0205288];
  M.~Fabinger and J.~McGreevy,
  ``On smooth time-dependent orbifolds and null singularities,''
  JHEP {\bf 0306}, 042 (2003)
  [arXiv:hep-th/0206196];
  H.~Liu, G.~W.~Moore and N.~Seiberg,
  ``Strings in time-dependent orbifolds,''
  JHEP {\bf 0210}, 031 (2002)
  [arXiv:hep-th/0206182].
}

\lref\nappiwitt{
  C.~R.~Nappi and E.~Witten,
  ``A Closed, expanding universe in string theory,''
  Phys.\ Lett.\ B {\bf 293}, 309 (1992)
  [arXiv:hep-th/9206078] and citations thereof.
}

\lref\Steve{
  L.~Fidkowski, V.~Hubeny, M.~Kleban and S.~Shenker,
  ``The black hole singularity in AdS/CFT,''
  JHEP {\bf 0402}, 014 (2004)
  [arXiv:hep-th/0306170];
  P.~Kraus, H.~Ooguri and S.~Shenker,
  ``Inside the horizon with AdS/CFT,''
  Phys.\ Rev.\ D {\bf 67}, 124022 (2003)
  [arXiv:hep-th/0212277].
}

\lref\HoriKT{
  K.~Hori and C.~Vafa,
``Mirror symmetry,''
  arXiv:hep-th/0002222.
}

\lref\BHfinalstate{
  G.~T.~Horowitz and J.~Maldacena,
``The black hole final state,''
  JHEP {\bf 0402}, 008 (2004)
  [arXiv:hep-th/0310281];
  D.~Gottesman and J.~Preskill,
 ``Comment on 'The black hole final state',''
  JHEP {\bf 0403}, 026 (2004)
  [arXiv:hep-th/0311269].
}

\lref\KT{
  J.~M.~Kosterlitz and D.~J.~Thouless,
``Ordering, Metastability And Phase Transitions In Two-Dimensional
Systems,''
  J.\ Phys.\ C {\bf 6}, 1181 (1973).
}
\lref\PolyakovRR{
  A.~M.~Polyakov,
``Interaction Of Goldstone Particles In Two-Dimensions. Applications To
Ferromagnets And Massive Yang-Mills Fields,''
  Phys.\ Lett.\ B {\bf 59}, 79 (1975).
}

\lref\NovikovAC{
  V.~A.~Novikov, M.~A.~Shifman, A.~I.~Vainshtein and V.~I.~Zakharov,
 ``Two-Dimensional Sigma Models: Modeling Nonperturbative Effects Of Quantum
 Chromodynamics,''
  Phys.\ Rept.\  {\bf 116}, 103 (1984).
}

\lref\FateevBV{
  V.~A.~Fateev and A.~B.~Zamolodchikov,
  ``Integrable perturbations of Z(N) parafermion models and O(3) sigma model,''
  Phys.\ Lett.\ B {\bf 271}, 91 (1991).
}

\lref\KlusonXN{
  J.~Kluson,
  ``The Schrodinger wave functional and closed string rolling tachyon,''
  Int.\ J.\ Mod.\ Phys.\ A {\bf 19}, 751 (2004)
  [arXiv:hep-th/0308023].
}

\lref\TseytlinMZ{
  A.~A.~Tseytlin,
``On The Structure Of The Renormalization Group Beta Functions In A Class Of
Two-Dimensional Models,''
  Phys.\ Lett.\ B {\bf 241}, 233 (1990).
}

\lref\SchmidhuberBV{
  C.~Schmidhuber and A.~A.~Tseytlin,
``On string cosmology and the RG flow in 2-d field theory,''
  Nucl.\ Phys.\ B {\bf 426}, 187 (1994)
  [arXiv:hep-th/9404180].
}

\lref\MosheXN{
  M.~Moshe and J.~Zinn-Justin,
``Quantum field theory in the large N limit: A review,''
  Phys.\ Rept.\  {\bf 385}, 69 (2003)
  [arXiv:hep-th/0306133].
}

\lref\BerkoozRE{
  M.~Berkooz, B.~Pioline and M.~Rozali,
``Closed strings in Misner space: Cosmological production of winding
strings,''
  JCAP {\bf 0408}, 004 (2004)
  [arXiv:hep-th/0405126].
}

\lref\CohenSM{
  A.~G.~Cohen, G.~W.~Moore, P.~Nelson and J.~Polchinski,
``An Off-Shell Propagator For String Theory,''
  Nucl.\ Phys.\ B {\bf 267}, 143 (1986).
}

\lref\SeibergHR{
  N.~Seiberg,
``From big crunch to big bang - is it possible?,''
  arXiv:hep-th/0201039.
}

\lref\ZZ{
  A.~B.~Zamolodchikov and A.~B.~Zamolodchikov,
``Liouville field theory on a pseudosphere,''
  arXiv:hep-th/0101152.
}

\lref\Hikida{
  Y.~Hikida and T.~Takayanagi,
``On solvable time-dependent model and rolling closed string tachyon,''
 Phys.\ Rev.\ D {\bf 70}, 126013 (2004)
  [arXiv:hep-th/0408124].
}

\Title{\vbox{\baselineskip12pt\hbox{hep-th/0506130} \hbox{SU-ITP-05/22}\hbox{SLAC-PUB-11283}}} {\vbox{ \centerline{
%
%
 The Tachyon at the End of the Universe
%
%
%
%
%
}}}
\bigskip
\bigskip
\centerline{John McGreevy and Eva Silverstein}
\bigskip
\centerline{{\it SLAC and Department of Physics, Stanford University, Stanford, CA 94305-4060}}
\bigskip
\bigskip
\noindent

We show that a tachyon condensate phase replaces the spacelike singularity in certain cosmological and black
hole spacetimes in string theory. We analyze explicitly a set of examples with flat spatial slices in various
dimensions which have a winding tachyon condensate, using worldsheet path integral methods from Liouville
theory.
In a vacuum with no
excitations above the tachyon background in the would-be singular region, we
analyze
the production of closed strings
in the
resulting state in the bulk of
spacetime.  We find a thermal result reminiscent of the Hartle-Hawking state, with tunably small energy density.
The amplitudes exhibit a self-consistent truncation of support to the
weakly-coupled
small-tachyon
region of spacetime.
We argue that the background is accordingly robust against back reaction, and that the
resulting string theory amplitudes are perturbatively finite, indicating
a resolution of the singularity and a mechanism to start or end time in string theory.  Finally, we discuss the
generalization of these methods to examples with positively curved spatial slices.

\bigskip
\Date{July 2005}

\newsec{Introduction}

Closed string tachyon condensation affects the dynamics of spacetime in interesting
and tractable ways
in many systems
\refs{\APS\othertach\TFA\CostaEJ-\garybubbles}.
%
In this paper,
we study circumstances in which closed string tachyon condensation plays a crucial
role in the dynamics of a system which has a spacelike singularity at the level of general relativity. The
singular region of the spacetime is replaced by a phase of tachyon condensate which lifts the closed string
degrees of freedom, effectively ending ordinary spacetime.

We will focus primarily on a simple
set of examples with shrinking circles,
in which we can make explicit
calculations exhibiting this effect.  Before specializing to this, let us start by explaining the relevant
structure of the stringy corrections to spacelike singularities appearing in a
more general context.
Much of this general discussion appeared
earlier in \Poly.\foot{The possibility of applying the worldsheet mass gap in higher dimensional generalizations
of \TFA\ was also independently suggested by A. Adams and M. Headrick.} Consider a general relativistic solution
approaching a curvature singularity in the past or future. The metric is of the form
\eqn\GRsing{ds^2=G_{\mu\nu}dx^\mu dx^\nu=-(dx^0)^2+R_i(x^0)^2 d\Omega_i^2+ ds^2_\perp}
with $R_i(x^0)\to 0$ for some $i$ at some finite time.  Here $\Omega_i$ describe spatial coordinates
 whose scale factor is varying in time
and $ds^2_\perp$ describes some transverse directions not directly
participating in the time dependent physics.

In the large radius regime where general relativity applies, the background \GRsing\ is described by a
worldsheet sigma model with action in conformal gauge
%
\eqn\wsGR{S_0\equiv
{1 \over 4 \pi \ap} \int d^2\sigma ~G_{\mu\nu}({X})\del_a{X}^\mu \del^a{ X}^\nu +{\rm fermions + ghosts}}
Here we are considering a type II or heterotic string with worldsheet supersymmetry
in order to avoid bulk tachyons.

As the space shrinks in the past or future,
at leading order in $\ap$ (\ie\ in GR) the corresponding sigma model kinetic terms for $\Omega$
develop small coefficients, leading to strong coupling
on the worldsheet.
This raises the possibility of divergent amplitudes
in the first quantized worldsheet path integral description from lack of suppression from the action.
This would correspond also to the development of
an effectively strong coupling in the spacetime theory as the size of the $\Omega$
directions shrink.

However, there is more to the story
in string theory.  Let us first consider the sigma model on
the angular geometry at fixed time $X^0$.
 When any of the radii $R_i $ in \GRsing\ is of order string scale, this strongly coupled sigma model
 is very different from the free flat-space theory.
 In particular, it can
dynamically generate a mass gap in the IR \refs{\PolyakovRR, \KT}.
In such cases the quantum effective action in this matter sigma model has terms of the form $\int d^2\sigma
{\cal O}_\Delta \Lambda^{2-\Delta}$ where $\Delta<2$ is the dimension of some relevant operator ${\cal O}$ and
$\Lambda$ is a mass scale. Hence the full string path integral \wsGR\ generates additional contributions to the
worldsheet effective action of the form
%
\eqn\wsTterm{S_T = -\int d^2\sigma  ~\mu f(X^0)~{\cal O}_\Delta ({X}_\perp, \Omega)
 +\int d^2\sigma \Phi(X^0)
R^{(2)} +{\rm fermions}}
where $f$ has dimension $2-\Delta$ in the unperturbed sigma model \wsGR.  We will henceforth refer to such
deformations as ``tachyons"; in the simplest case of an $S^1$ spatial component the corresponding mode is a
standard winding tachyon.  Let us discuss the big bang case for definiteness: the system becomes weakly curved
in the future (large positive $X^0$) and goes singular at some finite value of $X^0$ in the past. The
contribution \wsTterm\ goes to zero as $X^0\to +\infty$ since the sigma model is weakly coupled there.  So at
its onset the coefficient $f$ increases as $X^0$ decreases; \ie\ the effects of the term \wsTterm\ increase as
we go back in time in the direction of the would-be big bang singularity of the GR solution \GRsing. In the
simple case we will study in detail in \S2,3 below, a term of the form \wsTterm\ will arise from winding tachyon
condensation, and the operator $f$ will be of the form $e^{-\kappa X^0}$ for real positive $\kappa$ in the big
bang case.

This growth of \wsTterm\ as we approach the singularity contrasts to the suppression of the original sigma model
kinetic terms from the metric \wsGR.  In the Minkowski path integral, the
growth of the term \wsTterm\
serves to suppress fluctuations of the fields.  This provides a possibility of curing -- via
perturbative string effects -- the singular amplitudes predicted by a naive extrapolation of GR.  We will see
this occur explicitly in the examples we will study in detail in \S3.
\ifig\cloaking{
In string theory (with string length scale $l_s$), a tachyon condensate phase replaces
a spacelike singularity that would have been present at the level of general relativity.
}
{\epsfxsize3.5in\epsfbox{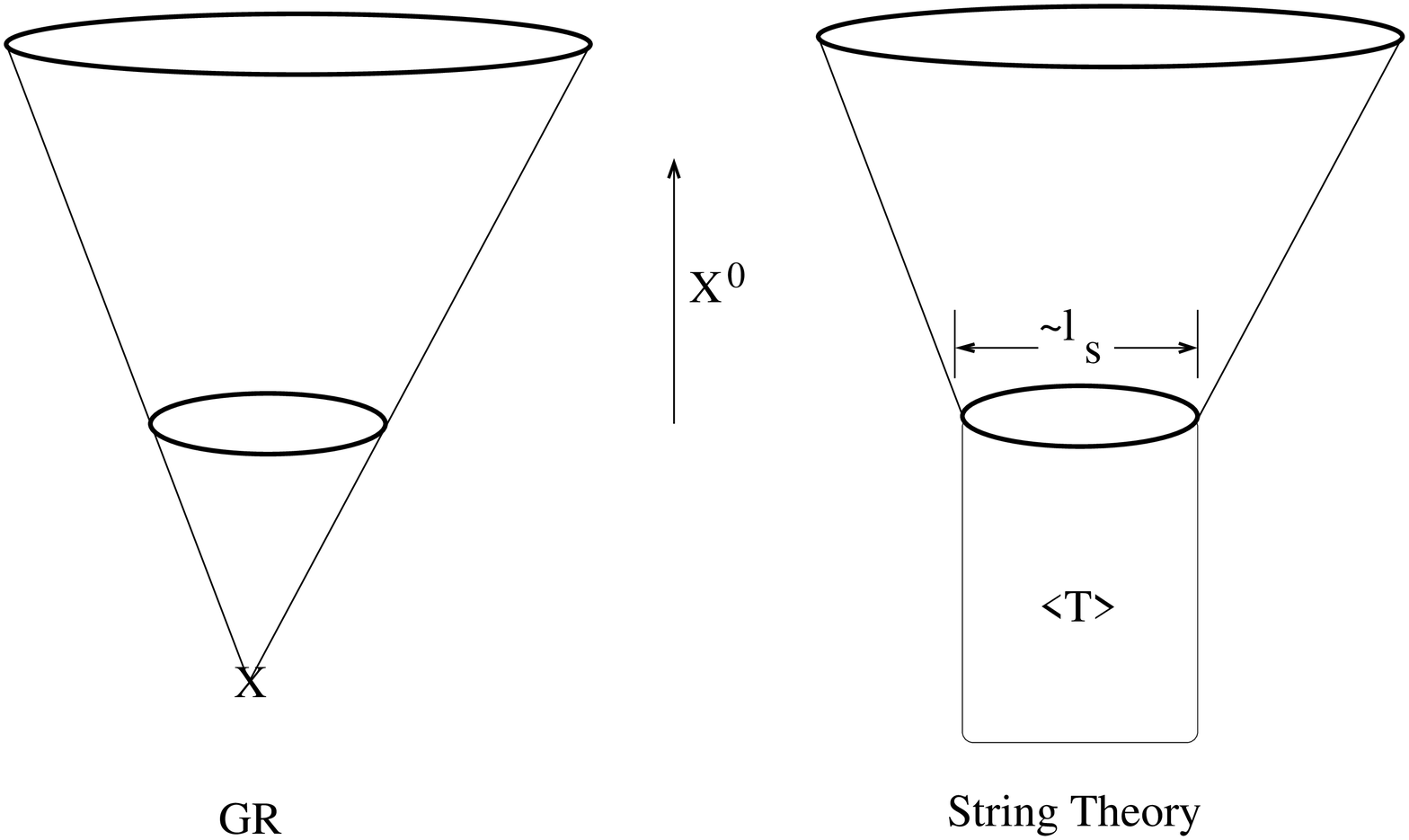}}

Because of the mass gap in the matter sector and the effect of the deformation $S_T$ on the spacetime mass
spectrum, the condensation of tachyons has long been heuristically argued to lift the string states and lead to
a phase of
``Nothing"\refs{\AtickWitten\KachruED\HarveyNA\DeAlwisPR\DeAlwisKP-\BarbonDD}.
In the examples \APS\ where conical singularities resolve into flat space, this is borne out in detail, as the
tip of the cone disappears in the region of tachyon condensation; a similar phenomenon was found for localized
winding tachyons in \TFA. In the present work, we will use methods from Liouville theory
(for a review see \refs{\seiberg,\GinspargIS,\teschner}).
We employ and
extend the methods of \refs{\TseytlinMZ\SchmidhuberBV\StromTak\DaCunhaFM\SchomerusVV\KlusonXN\Xiao-\Hikida}
 to perform systematic string theoretic calculations of amplitudes exhibiting
this effect in our temporally but not spatially localized case.
We study a particular vacuum state, analogous to the Euclidean vacuum.
In this vacuum the support of string theoretic
amplitudes is restricted to the bulk region of spacetime in a way that we can derive from the zero mode integral
of $X^0$ in the worldsheet path integral.

As discussed above, the metric coefficient $G_{\Omega\Omega}=R(X^0)^2$ in the worldsheet action $S_0+S_T$
\wsGR\wsTterm\ goes to zero at finite $X^0$. In the models we consider below we will set up the system such that
the approach to this singularity is parameterically slower than the timescale for the relevant term $S_T$ to
become important and lift the closed string degrees of freedom.  This will avoid strong dilaton effects as well
as effects of the shrinking space, and provides a mechanism which is inherently perturbative in $g_s$.
The amplitudes will exhibit limited support in the spacetime, contributing only in the bulk region away from the
epoch when the couplings become important.  This is similar to the situation in spacelike Liouville theory,
where similar strong coupling effects are avoided by the presence of the Liouville wall, and similar
computational methods apply, though the physical mechanism for suppressing amplitudes is different in the two
cases. Hence a self-consistent perturbative analysis is available. Relatedly, black hole formation of the sort
found in \refs{\GaryJoeetal,\LMS} is evaded here: the tachyon lifts the degrees of freedom of the system before
the Planckian regime is reached.

The observables of this theory are correlation functions of integrated vertex operators computed by the
worldsheet path integral with semiclassical action $S_0+S_T$ \wsGR\wsTterm; let us now discuss their spacetime
interpretation.
As in Liouville theory, the form of these operators is known in the weakly curved bulk region
where there is no tachyon condensate ($X^0\to\infty$ in the big bang case); there they asymptote in locally flat
coordinates to operators of the form
\eqn\Vopsgen{V_{\vec k, n}\to e^{i\vec k\cdot \vec X}e^{i\omega(\vec k, n) X^0} \hat V_n ~~ {\rm as} ~~
X^0\to\infty}
where $n$ labels the string state with mass $m_n$ coming from oscillator excitations created by $\hat V_n$,
$\vec k$ its spatial momentum, and $\omega^2=\vec k^2+m_n^2$.\foot{Although we do not know the form of these
operators in the regime where the corrections \wsTterm\ become important, we do know their conformal dimensions
by virtue of their form in the bulk region of the spacetime.  This is as in Liouville theory, where one knows
the operators and the stress tensor away from the Liouville wall, and hence the spectrum of dimensions. And as
in Liouville theory, an important question which we will address
is where the amplitudes built from these
operators have their support.}
Integrated correlation functions of these operators have the interpretation
as components
of the state of the strings in
the bulk region of spacetime
in a basis of multiple free string modes.
In our example below, we will focus on a vacuum with no excitations above
the tachyon condensate
in the would-be singular region, and compute
the resulting state of
perturbative strings in the bulk region.  This is
a string-theoretic analogue of the
Hartle-Hawking State (equivalently, the Euclidean Vacuum) on our time-dependent background.


We will treat the condensing tachyon in string perturbation theory.
We obtain a self-consistent analysis
at weak string coupling,
in systems
with bulk supersymmetry, and with supersymmetry breaking near the would-be singular region 
similar to that
expected in the early universe and inside black holes. Other interesting recent
work on perturbative closed string
calculations in time dependent backgrounds includes \refs{\LMS,\nappiwitt}.  In our case the tachyon
condensation, related to the supersymmetry breaking of the time dependent background, plays a crucial role, in a
way anticipated in \Poly.

It would be interesting to relate our analysis to other approaches based on non-perturbative formulations of the
theory \refs{\Steve\Gary-\Matrixcosmo}.  These approaches may provide a complete nonperturbative dual
formulation of observables in spacetimes with singularities at the level of GR.  On the other hand, the
dictionary between the two sides is sometimes rather indirect as applied to approximately local processes on the
gravity side.  A useful feature of the current approach is that the tachyon condensation provides a direct
gravity-side mechanism for quelling the singularity.  It would be interesting to see how this information is
encoded in the various dual descriptions.

\lref\BartonI{
  H.~Yang and B.~Zwiebach,
  ``A Closed String Tachyon Vacuum?,''
  arXiv:hep-th/0506077.
}

\lref\SenNF{
  A.~Sen,
  ``Tachyon dynamics in open string theory,''
  arXiv:hep-th/0410103.
}

Finally,
analogously to the case of open string tachyons
(for a review, see \SenNF),
closed string tachyons may be a subject well-studied via closed string field theory; a candidate
``nothing" state obtained from bosonic closed string bulk tachyon condensation was recently
presented in \BartonI. It is clear (as we will review as we go) that the physics of the tachyon condensate is
stringy -- low energy effective field theory is not sufficient. In the setup we consider here, perturbative
methods using techniques from Liouville theory will suffice, but in more general situations the off shell
methods of string field theory may be required.

In the next two sections we set up and analyze a class of realizations of the mechanism.
In \S4 we describe the generalization to positive spatial curvature,
which is velocity-dominated.
Philosophy-dominated comments are restricted to the concluding section.

\newsec{Examples with winding tachyons}

In this section, we will introduce the simplest backgrounds we will study;
those with flat spatial slices which
expand at a tunable rate.  We will start with an example pertaining to 2+1
dimensional black holes (reducing to
the 1+1 dimensional Milne spacetime inside), and then generalize to higher
dimensional flat FRW cosmology with
topologically nontrivial spatial slices and radiation.

\subsec{The Milne Spacetime}

Consider the Milne spacetime described by the metric
\eqn\milne{ds^2= -(dx^0)^2+v^2 (x^0)^2 d\Omega^2 + d\vec x^2}
For $x^0>0$, this solution describes a growing $S^1$ along the $\Omega$ direction. At $x^0=0$ there is a
spacelike big bang singularity, and general relativity breaks down.  The evolution from $x^0=-\infty$
to $x^0=0$ similarly describes an evolution toward a big crunch singularity. This geometry appears inside
$2+1$ dimensional black holes, BTZ black holes in $AdS_3$.\foot{``Whisker" regions with closed timelike curves
also appear in the maximally extended spacetime; our methods here will also have the effect of excising these
regions, as obtained in other examples in \CostaEJ.} We will show that for a wide class of string theories, the
spacelike big bang or big crunch singularity \milne\ is evaded -- the regime $|vx^0|<l_s$ is replaced by a phase
of tachyon condensate.

In particular, to avoid bulk tachyons, consider type II, type I or
heterotic string theory on the spacetime \milne. Take antiperiodic
boundary conditions around the $\Omega$ circle for spacetime fermions. Further consider the regime of parameters
where $v\ll 1$.  In addition to providing the control we will require, the last two conditions correspond to those
appropriate for small BTZ black holes which can form naturally from excitations in pure $AdS_3$ (which has
antiperiodic boundary conditions for fermions around the contractible spatial circle surrounding the origin).

With these specifications, we can determine with control the spectrum of string theory on the spacetime \milne\
for $x^0\ne 0$.   In the regime
\eqn\tachcond{v^2(x^0)^2 \le l_s^2}
a closed string winding mode becomes tachyonic and hence important to the dynamics.  The regime $v|x^0|\le l_s$
of the singularity in \milne\ is replaced by a phase of tachyon condensate.  This offers a concrete avenue
toward resolving a spacelike singularity in string theory, and a corresponding notion of how time can begin or
end.

This in itself is worth emphasizing.  The problem of bulk tachyon condensation is often motivated by the
question of the vacuum structure of string theory.  The present considerations provide an independent motivation
for pursuing the physics of closed string tachyon condensation: it appears crucially in a string-corrected
spacelike singularity.  In our system here there is no tachyonic mode in the bulk of spacetime: for a
semiinfinite range of time the system is perturbatively stable.  That is, the tachyon phase is localized in
time.  As we will see, this provides significant control over the problem even though the condensation is not
also localized in space.

\subsec{Flat FRW with topology}

Next let us set up a somewhat more realistic case which shares the essential features of the above example.
Consider flat-sliced FRW cosmology with bulk metric
\eqn\FRWrad{ds^2=-(dx^0)^2+a^2(x^0)d\vec x^2+ds^2_\perp}
with $\vec x$ a 3-dimensional spatial vector and $ds^2_\perp$ describing the extra dimensions. Let us consider
some periodicity in the spatial directions $\vec x$:  $\vec x\equiv \vec x+\vec L_I$; \eg\ letting $I$ run from
1 to 3 produces a spatial torus (for simplicity let us take a rectangular torus).  In real cosmology, such
topology could well exist at sufficiently large scales (most generically well outside our horizon today due to
inflation), but if present would play a role in the far past in the epoch of the would-be big bang singularity.
%
\lref\LindeNZ{
  A.~Linde,
  ``Creation of a compact topologically nontrivial inflationary universe,''
  JCAP {\bf 0410}, 004 (2004)
  [arXiv:hep-th/0408164].
}
(See \eg\ \LindeNZ\ for one recent discussion of spatial topology.).

Let us study the above system in the presence of a
stress-energy source.
For definiteness, consider a homogeneous bath of
radiation.
The Friedmann equation implies
%
\eqn\radscale{a(x_0)=a_0\sqrt{x^0-t_0}}
where the coefficient $a_0$ can be tuned by dialing the amount of radiation.

In particular, as in the above example \tachcond, we can choose the radiation density and hence $a_0$ so as to
obtain a slow expansion of the toroidal radii $R\sim L a(x^0)$ as the
smallest radius
passes through the string scale.
Again considering antiperiodic boundary conditions for fermions along one or more of the 1-cycles of the torus,
we then obtain in a controlled way a winding tachyon in the system as the radius $R\sim L a(x^0)$ of a circle
passes below the string scale. The would-be big bang singularity is again replaced by a tachyon condensate
phase, whose consequences we will analyze in detail in the next section.

\newsec{Examples with winding tachyons:  some basic computations of observables}

In this section, we develop a systematic computational scheme to compute physical observables in this system,
assess back reaction, and test and make more precise the proposition that tachyon condensation lifts closed
string excitations  (leading to a phase we will refer to as a Nothing state).

\subsec{Wick rotation}

Let us start by defining the path integral via appropriate Wick rotation.  In its original Lorentzian signature
form, the tachyon term appears to increase the oscillations of the integrand, hence suppressing contributions in
the region of the tachyon condensate.  As is standard in quantum field theory, we will perform a Wick rotation
to render the path integral manifestly convergent (up to, as we will see, divergences at exceptional momenta
expected from the bulk S-matrix point of view). The path integral in conformal gauge includes an integral over
the target space time variable $X^0$, which has a negative kinetic term in the worldsheet theory. Because this
field also appears necessarily in the tachyon interaction term (which is proportional to $e^{-\kappa X^0}$,
specializing to the big bang case), we will find it convenient to Wick rotate the worldsheet theory to directly
obtain
real positive
kinetic terms for $X^0$ without rotating the contour for $X^0$ integration; this
will entail rotating the contours for the spatial target space coordinates as well as continuing $\mu$ in a way
we will specify. (Alternatively one could rotate $X^0$ as is standardly done in the free theory, and continue in
$\kappa$ at the same time, as in \StromTak.)

\subsubsec{Prelude:  worldline quantum field theory}

Before turning to the full string path integral, let us briefly describe a much simpler analogue of our system
which arises in the worldline description of quantum field theory, as emphasized in \Strompart. Consider a
relativistic particle action
\eqn\partac{ S= \int d\tau \biggl(-(\partial_\tau X^0)^2+(\del_\tau\vec X)^2-(m_0^2+\mu^2 e^{-2\kappa
X^0})\biggr)}
where we have included a time-dependent mass squared term $m^2(X^0)=m_0^2+\mu^2 e^{-2\kappa X^0}$.

For $\mu^2>0$, this theory describes a
particle with a time-dependent positive mass-squared that
increases exponentially in the past $X^0\to -\infty$.  The potential term in the relativistic worldline action
leads to a lifting of particles in the region where it becomes important.  If one starts with none of these
massive modes excited in the past, then the future state gets populated due to the time dependent mass. The
Bogoliubov coefficient $\beta_{\vec k}$ describing mixing of positive and negative frequency modes has magnitude
$e^{-\pi\omega/\kappa}$ with $\omega=\sqrt{\vec k^2+m_0^2}$ the frequency of the particle modes in the region
$X^0\to +\infty$.  We will find similar features in our string theoretic examples, where the phase in which
states are lifted replaces a spacelike singularity.

For $\mu^2<0$, this theory describes a system with time-dependent negative mass squared.
A particle with
positive mass squared in the far future becomes tachyonic in the far past.  One could formally start again with
no excitations in the past, but this would be unnatural as the tachyonic modes there would condense.

In Lorentzian signature, the worldline path integral is
\eqn\partPI{\int [dX] ~e^{iS}}
If we continue $\tau\equiv e^{i\gamma} \tau_\gamma$ and $\vec X\equiv e^{i\gamma}\vec X_\gamma$, taking $\gamma$
continuously from $0$ to $\pi/2$, we obtain a path integral
\eqn\Euclpart{\int [dX_E]~\exp{\left[-\int d\tau_E
\left( (\del_{\tau_E}X^0)^2+(\del_{\tau_E}\vec X)^2+\mu^2 e^{2\kappa X^0}
\right) \right]}  ~~.
}
Ambiguities in defining the $X^0$ integral
correspond to choices of vacuum state.
In order to obtain a convergent path integral, we can continue $\mu^2\to -\mu^2$ (\ie\ $\mu\to e^{-i\gamma} \mu$) as
we do the above Wick and contour rotation, compute the amplitudes, and then continue back.  That is, our
computation is related to one in a purely spacelike target space via a reflection of the potential term in the
worldline theory.

This reflection appears also in the direct spacetime analysis of particle production in field theory with time
dependent mass. The Heisenberg equation of motion satisfied by the Heisenberg picture fields in spacetime takes
the form of a Schr\"odinger problem for each $\vec k$ mode.  The effective potential in the Schr\"odinger
problem is $U_{eff}=-(m^2(X^0)+\vec k^2)$. This leads to highly oscillating mode solutions as $X^0\to -\infty$,
reflecting the exponentially increasing mass in the far past.

With this warmup, let us now turn to the case of string theory with a tachyon condensate, which has a worldsheet
potential term analogous to the mass squared term in the field theory case.
We will study both the heterotic and type II theories on
our background.

\subsubsec{Heterotic Theory}

In the RNS description of the heterotic string, the worldsheet theory has local (0,1) supersymmetry.
This case is in some ways the simplest
for studying closed string tachyons using the string worldsheet description -- unlike the bosonic theory, there is
no tachyon in the bulk region where the $S^1$ is large; unlike the type II theory the worldsheet bosonic
potential is automatically nonnegative classically (as in the open superstring theory).

There is a choice of discrete torsion in the heterotic theory on a space with shrinking Scherk-Schwarz circle.
The background can be regarded as a $\IZ_2$ orbifold of a circle by a shift halfway around combined with an
action of $(-1)^F$ where $F$ is the spacetime fermion number.  Combining this $\IZ_2$ with that of the
left-moving fermions (in say the SO(32) Heterotic theory) yields two independent choices of action of the left
moving GSO on the states of the Scherk-Schwarz twisted sector. A standard choice,
giving rise to a Hagedorn tachyon,
is to act trivially on the Scherk-Schwarz twisted sector; this yields a twisted sector tachyon made
from momentum and winding modes \AtickWitten.

This would also be the most natural choice for us in some sense, since the usual spacelike singularities in cosmology
and inside black holes are a priori independent of Yang-Mills degrees of freedom.  However because the
winding+momentum twisted tachyon in the above case is a nonlocal operator on the worldsheet in both T-duality
frames, we will make here a technically simpler choice.  Namely, we can choose the discrete torsion such that
the left-moving GSO projection acts nontrivially on the states of the twisted sector, yielding a twisted tachyon
created by a left moving fermion and a winding operator.

In the heterotic theory we have target space coordinates given by (0,1) scalar superfields ${\cal
X}^\mu=X^\mu+\theta^+\psi_+^\mu$ and left moving fermion superfields $\Psi_-^a= \psi_-^a+\theta^+F^a$ containing
auxiliary fields $F^a$.  In terms of these fields we have a Lorentzian signature path integral
\eqn\LorPIHet{G(\{ V_n\})\equiv \int [d{\cal X}][d\Psi_-][d({\rm ghosts})]d({\rm moduli})~e^{iS}\prod_n \biggl(i \int d\sigma
d\tau V_n[\cal X]\biggr) }
where the semiclassical action is
\eqn\superacHet{\eqalign{ iS= & i \int d\sigma d\tau d\theta^+ \biggl( D_{\theta^+}{\cal X^\mu}\del_-{\cal
X^\nu}G_{\mu\nu}({\cal X}) -\mu \Psi_-: e^{-\kappa{\cal X}^0}\cos(w\tilde\Omega) : \cr &
+\Psi_-^aD_{\theta^+}\Psi_-^a + ({\rm dilaton}) \biggr) +iS_{{\rm ghost}}\cr }}
and $V_n[{\cal X}]$ are vertex operator insertions.  Here $\tilde\Omega$ is the T-dual of the coordinate
$\Omega$ on the smallest circle in the space (let us consider for genericity a rectangular torus, whose smallest
cycle will play a leading role in the dynamics); $\cos w\tilde\Omega$ is the winding operator for strings wrapped
around the $\Omega$ direction.

The case of no insertions corresponds to the vacuum amplitude $Z$. The fluctuations of the worldsheet fields in
\LorPIHet\ generate corrections to the action \superacHet; for example, the term proportional to $\mu$ coming
from the tachyon condensate is marginal but not exactly marginal. This is similar to the form of the Liouville
wall in Liouville field theory, which is a priori only semiclassically given by a pure exponential.

Similarly, the form of the vertex operators is known semiclassically.  Because the bulk region of the geometry
\milne\ is approximately flat space, we may identify the $V_n$ with operators of the form
\eqn\superVHet{V_{\vec k, n}\to e^{i\vec k\cdot \vec{\cal X}}e^{i\omega(\vec k, n) {\cal X}^0} \hat V_n ~~ {\rm
as} ~~ X^0\to\infty}
where as in \Vopsgen\ we have pulled out the oscillator and ghost contributions into $\hat V$.

Finally, at the semiclassical level the dilaton is also known:  it goes to a constant
\eqn\dilsemi{\Phi\to \Phi_0 ~~ {\rm as} ~~ X^0\to +\infty ~.}
In particular, the tachyon vertex operator in \superacHet\ is semiclassically marginal without an additional
dilaton contribution (though not exactly marginal) and the metric terms solve Einstein's equations.  The path
integral over fluctuations of the fields will generate corrections to these semiclassical statements
\superacHet\superVHet\dilsemi.

The basic physical effect of the tachyon condensate is the oscillations it induces in the path integral,
suggesting a suppression of contributions rather than a divergence in the would-be crunch region.  For some
computations, it is useful to Wick rotate in evaluating the path integral (though as we will see below some
computations such as the 1-loop partition function could be evaluated more directly).

Let us Wick rotate the worldsheet time coordinate $\tau$, the spatial target space coordinates $\vec {\cal
X}(\sigma, \tau)$ (including $\tilde\Omega$), and the parameters $\mu$ and $\vec k$ by
%
\eqn\rotationsHet{\tau \equiv e^{i\gamma}\tau_\gamma ~~~ \vec {\cal X}\equiv e^{i\gamma} \vec {\cal X}_\gamma
~~~
\mu= e^{-i\gamma} \mu_\gamma ~~~ \vec k= e^{-i\gamma}\vec
k_\gamma}
where $\gamma$ is a phase which we will rotate from $0$ to $\pi/2$.  This produces a Euclidean path integral for
the worldsheet theory (where we label the quantities rotated to $\gamma=\pi/2$ by a subscript $E$)
\eqn\EuclPIHet{G(\{ V_n\})\equiv \int [d\vec{\cal X}_E] [d{\cal X}^0][d\Psi_-][d({\rm ghosts})]d({\rm
moduli})~e^{-S_E}\prod_n \int (-1) d\sigma d\tau_E V_{n, -i\vec k_E}[{\cal X}^0,i\vec {\cal X}_E]}
with Euclidean action
\eqn\EuclacHet{\eqalign{ S_E= & \int d\sigma d\tau_E d\theta^+\biggl(D_{\theta^+}{\cal X}^0\del_-{\cal
X}^0 +v^2({\cal X}^0)^2D_{\theta^+}\tilde\Omega_E\del_-\tilde\Omega_E + G_{ij} D_{\theta^+}{\cal X}^i_{\perp,E}
\del_-{\cal X}^j_{\perp,E} \cr & -i\mu_E e^{-\kappa {\cal X}^0}\cosh(w\tilde\Omega_E)\cr & +
\Psi_-^aD_{\theta^+}\Psi_-^a +({\rm dilaton})\biggr)+S_E({\rm ghost})\cr }}
Here $\vec{\cal X}_E\equiv (\Omega_E, \vec{\cal X}_{\perp E} )$ refers to the worldsheet superfields corresponding
to the spatial target space coordinates, and we have plugged in the spacetime metric \milne.

The procedure \rotationsHet\ alone yields a formal path integral whose tachyon term is not marginal in the
Euclidean theory.  In order to retain semiclassical conformal invariance, we could include in \rotationsHet\ a
linear dilaton coefficient:  $Q_\gamma\equiv \gamma (2/\pi) Q$ where $Q$ corresponds to the coefficient of the
linear dilaton which renders the tachyon vertex operator marginal at leading order.  This is similar to part of
the procedure followed in \eg\ \StromTak.  Alternatively we can analyze the theory for $Q=0$, in which case the
intermediate steps in the analysis involve path integral computations in a nonconformal Euclidean worldsheet
theory, obtained by analytic continuation in $\kappa$ from a conformal Euclidean theory.  As we will see, for
our computations of the vacuum energy and of particle production, both procedures yield the same result.

In the specification of operators in the bulk region we have neglected terms proportional to
 the slow velocity $v$ by which the circles
shrink in the bulk metric \FRWrad\milne. Relatedly, $\mu$ could depend weakly on the other spatial directions
$\vec X_\perp$; we will ignore this for the purposes of the current discussion though it is simple to
incorporate.

In fact the small velocity approximation will play an important role more generally in our analysis of the
approach to the singularity. As it stands, the path integral \EuclPIHet\EuclacHet\ does not extend over all
values of $X^0$: the metric term $G_{\Omega\Omega}$ of classical GR goes to zero in finite $X^0$ in the past. As
we will see, for a constant radius circle of size $L$, a winding tachyon condensate will produce a truncation of
the support of amplitudes to a range of $X^0$ of order $(\ln \mu)/\kappa$ in the region of the condensate.
 We can
arrange the parameters in our worldsheet CFT such that the velocity is sufficiently small that this range of
time is far smaller than the time it takes to reach the singularity starting from a circle of size $L$.
The basic idea is that
the effective Newton constant does get large as the
space shrinks,
but the effects of the tachyon kick in first.
Namely,
consider a winding tachyon which turns on when the circle size is $L$.  The corresponding value of $\kappa$ is
$\kappa=\sqrt{1-(L/l_s)^2}$.  Set $v$ such that
\eqn\rangechoice{{L\over v}\gg -{ {\ln\mu}\over\kappa}}
This specification, combined with the $(\ln\mu)/\kappa$ truncation of the amplitudes' support in the $X^0$
direction to be derived below, yields a self-consistent perturbative string analysis. Note that the worldsheet
potential term in the heterotic theory is classically always non-negative.

From this well-defined path integral \EuclPIHet\EuclacHet\ we will obtain the $\mu$ dependence of amplitudes
using methods developed for Liouville theory which also apply to our theory.  This will enable us to read off
the effect of the tachyon on the support of amplitudes, and will determine the spectrum of particles produced
due to the time dependence of the tachyon background.

\subsubsec{Type II Theory}

Let us next briefly include the type II version of the above formulas. In the type II theory, we have (1,1)
scalar superfields ${\cal X}^\mu=X^\mu+\theta^+\psi_+^\mu+\theta^-\psi_-^\mu+\theta^+\theta^-F^\mu$. In terms of these, we
have a Lorentzian signature path integral
\eqn\LorPI{G(\{ V_n\})\equiv \int [d{\cal X}][d({\rm ghosts})]d({\rm moduli})~e^{iS}\prod_n \biggl(i \int d\sigma d\tau
V_n[\cal X]\biggr) }
where the semiclassical action is
\eqn\superac{\eqalign{iS= & i \int d\sigma d\tau d\theta^+ d\theta^- \biggl( D_{\theta^+}{\cal
X^\mu}D_{\theta^-}{\cal X^\nu}G_{\mu\nu}({\cal X})  -\mu
: e^{-\kappa{\cal X}^0}\cos(w\tilde\Omega) : \cr &
+({\rm dilaton}) \biggr)+iS_{{\rm ghost}}}}
and $V_n[{\cal X}]$ are vertex operator insertions.
As in the heterotic case, the form of the vertex operators is known
in the flat space region to be of the form \superVHet.
The dilaton is \dilsemi.
%
%
%
%
%

Let us Wick rotate the worldsheet time coordinate $\tau$, the spatial target space coordinates $\vec X(\sigma,
\tau)$ (including $\tilde\Omega$), and the parameters $\mu$ and $\vec k$ by
\eqn\rotations{\tau \equiv e^{i\gamma}\tau_\gamma ~~~ \vec {\cal X}\equiv e^{i\gamma} \vec {\cal X}_\gamma ~~~
\mu= e^{-i\gamma} \mu_\gamma ~~~ \vec k= e^{-i\gamma}\vec
k_\gamma}
where $\gamma$ is a phase which we will rotate from $0$ to $\pi/2$.  This produces a Euclidean path integral for
the worldsheet theory (where we label the quantities rotated to $\gamma=\pi/2$ by a subscript $E$)
\eqn\EuclPI{G(\{ V_n\})\equiv \int [d\vec{\cal X}_E] [d{\cal X}^0][d({\rm ghosts})]d({\rm
moduli})~e^{-S_E}\prod_n \int (-) d\sigma d\tau_E V_{n, -i\vec k_E}[{\cal X}^0,i\vec {\cal X}_E]}
with Euclidean action
\eqn\Euclac{\eqalign{S_E= & \int d\sigma d\tau_E d\theta^+ d\theta^-\biggl(D_{\theta^+}{\cal
X}^0D_{\theta^-}{\cal X}^0 + v^2({\cal X}^0)^2D_{\theta^+}\tilde\Omega D_{\theta^-}\tilde\Omega + G_{ij}
D_{\theta^+}{\cal X}^i_{\perp,E} D_{\theta^-}{\cal X}^j_{\perp,E} \cr & -i\mu_E e^{-\kappa {\cal
X}^0}\cosh(w\tilde\Omega_E)+({\rm dilaton})\biggr)+S_E({\rm ghost}) ~.}  }

Again, as discussed below \EuclacHet, we could include a $\gamma$-dependent shift of the linear dilaton
coefficient (which vanishes in our physical critical-dimension Lorentzian bulk theory) if we wish to maintain
semiclassical conformal invariance in the intermediate (rotated) steps of the calculations.

\subsec{Vacuum Amplitude and Back Reaction}

In this subsection we will present computations exhibiting the effect we advertised above that the amplitudes
will be limited in their support to the weakly-coupled bulk.
%
Let us start with the vacuum amplitude.  At one loop, this is defined by the amplitude \EuclPI\ with no vertex
operator insertions, evaluated on a genus one worldsheet; let us call this amplitude $Z_1$.
In a time
dependent background,
one must specify the vacuum in which the amplitudes are defined (for example, one
definition of the S matrix would involve in-vacuum to out-vacuum amplitudes).  We will return to the question of
the vacuum after computing the first quantized path integral defined above at this 1-loop order.

In the bulk, this quantity describes a trace over spacetime single-particle states.  In our case, the integral
over the zero mode of $X^0$ will work differently than in flat space, and we will determine from this the
support of the amplitudes as well as the quantum corrections to the stress energy in spacetime. In particular,
as in Liouville field theory, we will find this amplitude to be supported only in the bulk region where the
tachyon condensate is small.  This supports the interpretation of the tachyon condensate as lifting the closed
string degrees of freedom.  Further, with our asymptotic supersymmetry in the bulk region this also provides a
useful bound on the back reaction in the model.  Finally, the imaginary part of the amplitude will provide
information about the vacuum with respect to which the amplitude is being computed from the spacetime point of
view.

Following \refs{\LFTpartition,\seiberg}, let us compute first the quantity $\partial Z_1/\partial\mu$ and perform the path
integral by doing the integral over the $X^0$ zero modes first.  That is, decompose
\eqn\splitmodes{X^0\equiv X^0_0 + \hat X^0(\sigma, \tau_E)}
where $\hat X^0$ contains the nonzero mode dependence on the worldsheet coordinates $\sigma, \tau_E$.\foot{The
reader should be grateful that we are suppressing the atomic number and baryon number indices on $ {}_0^0 X^0_0
$. }
The path integral measure then decomposes as $[dX^0]=dX^0_0 [d\hat X^0]$.  We obtain for heterotic and type II
respectively
\eqn\HetpartI{\eqalign{{\del Z_1^{(Het)}\over{\del\mu_E}}=\int [d\vec{\cal X}_E][d\Psi_-] & [d({\rm ghosts})]
d({\rm moduli})[d\hat {\cal X}^0]dX^0_0 \cr & \biggl(-\int d\sigma d\tau_E d\theta^+ \Psi_- e^{-\kappa {\cal
X}^0}i \cosh(w\tilde\Omega_E)\biggr) e^{-S_E} \cr }}
\eqn\partI{\eqalign{{\del Z_1^{(II)}\over{\del\mu_E}}=\int [d\vec{\cal X}_E] & [d({\rm ghosts})]  d({\rm
moduli})[d\hat {\cal X}^0]dX^0_0 \cr & \biggl(-\int d\sigma d\tau_E d\theta^+ d\theta^- e^{-\kappa {\cal X}^0}i
\cosh(w\tilde\Omega_E)\biggr) e^{-S_E} \cr }}
Decomposing $e^{-\kappa {\cal X}^0}=e^{-\kappa X^0_0} e^{-\kappa \hat{\cal X}^0}$, we can change variables in
the zero mode integral to $y\equiv e^{-\kappa X^0_0}$ and integrate from $y=0$ to $y=\infty$ as $X_0^0$ ranges
from $\infty$ to $-\infty$\foot{
Note that the support of the integrand is negligible in the
added region $X_0^0 \in [ -\infty, 0]$.
}. For each point in worldsheet field space, the zero mode integral is of the form
\eqn\yform{\int_0^\infty dy ~e^{-Cy}={1\over C}}
where the coefficient $C$ is the nonzeromode part of the tachyon vertex operator in $S_E$, integrated over
worldsheet superspace.

Integrating over $\theta^\pm$ produces a worldsheet potential term contributing to $C$. For regions of field
space where $C$ is positive, the integral \yform\ converges.    For regions of negative $C$ the equation \yform\
gives {\it a} formal definition of the function by analytic continuation. However, it is important to keep in
mind the physical distinction between these two cases. As discussed above in the quantum field theory case, when
$C$ is positive this corresponds to a time dependent massing up of modes, while negative $C$ corresponds to time
dependent tachyonic masses.  In the latter case, the formal definition \yform\ describes an analytic
continuation of an interesting physical IR divergence.

In the heterotic theory, this coefficient $C$ is nonnegative everywhere in field space for $\mu^2>0$, at least
classically.  Hence the computation \yform\ applies directly.

In the type II theory, this coefficient can become negative near particular points in $\tilde\Omega$ and $\vec
X_\perp$. In regions where the potential is positive, \yform\ applies, and as we will see leads to a truncation
of the support of the closed string states.  However, in regions where $C$ is negative, there are physical
instabilities remaining.  These localized instabilities we interpret as subcritical type 0 tachyons.  In
particular, in \S4.1\ we will see using linear sigma model techniques that the GSO projection acts on the
corresponding subcritical theory as in type 0.

This analysis yields
\eqn\Zderiv{{\del Z_1\over{\del\mu_E}}=-{1\over{\kappa\mu_E}}\hat Z_1}
where $\hat Z_1$ is the partition function in the free theory (with no tachyon term in the action) and with no
integral over the zero mode of $X^0$.  Referring to the functional measure for the rest of the modes (including
all fields) as $[d({\rm fields})]'$ this is
\eqn\restZ{\hat Z_1=\int [d({\rm fields})]'[d({\rm ghosts})]d({\rm moduli})~e^{-\hat S_E}}
where $\hat S_E$ is the action (\EuclacHet\ and \Euclac\ respectively for heterotic and type II) with $\mu=0$.
%
%
%
Finally, integrating with respect to $\mu$ yields the result for the
1-loop partition function
\eqn\Zresult{Z_1=-{\ln(\mu_E/\mu_*)\over\kappa}\hat Z_1.}
Here $\mu_* = e^{\kappa X^0_*} $ where $X^0_*$ is an IR cutoff on the $X^0$ field space in the free-field
region. As discussed above, this is valid for regions of the worldsheet field space where the potential is
positive, which is always true in the Heterotic case and true for most contributions in type II.

It is worth remarking here that this result follows simply from the exponential form of the potential in the
tachyon vertex operator.  The above derivation formally goes through
whether rotated back to target space Minkowski signature or not,
(since the zeromode does not appear in the kinetic term),
reducing the amplitude to a tachyon-free
computation $\hat Z$.  Relatedly it is insensitive to the issue of rotating in $Q$ as discussed below
\EuclacHet.

To interpret this result, recall that in a background of $d$-dimensional flat space, the partition function scales
like the volume of spacetime:  integration over the zero modes of the $X^\mu$ fields yields the factor
$\delta^{(d)}(0)=V_d=V_0V_{d-1}$ where $V_{d-1}$ is the volume of space and $V_0$ is the volume of the time
direction.  In our present case \Zresult, the spatial extensivity reflected in the factor $V_{d-1}$ is still
present. But the volume of time $V_0$ has been truncated to $-{1\over \kappa}\ln \mu/\mu_*$.  This corresponds
to the range of $X^0$ where the tachyon is absent.  Again, this is similar to the situation in spacelike
Liouville field theory \LFTpartition, where the Liouville wall cuts off the support of the partition function.

This result has several implications.  First, it provides a concrete verification that the string states are
lifted in the tachyon phase, for positive worldsheet potential, supporting the interpretation of this phase as a
theory of Nothing.
Combined with the specification \rangechoice\ this result justifies the
use of the worldsheet path integral with metric coefficients going to zero in finite time, as the amplitudes are
not supported in this region.  Note that in particular all states are lifted -- the would-be tachyon and graviton
fluctuations are absent and hence back reaction is suppressed.
As emphasized above, we are computing
in a particular state analogous to the Euclidean or Hartle-Hawking vacuum, with no excitations in
the far past.  Putative vacua with excitations in the far past will be analyzed separately \usnext.  The
analagous context of quantum field theory with exponentially growing mass again suggests that the string states
will become lifted (effectively go to infinite mass) in an appropriate sense; moreover an interesting BRST
anomaly appears which may limit the consistency of such vacua in the string theory context. (See
\Schomerusanomaly\ for some discussion of the issue in field theory).

Second, it indicates that the 1-loop vacuum energy is only supported in the bulk region of the spacetime.
Because the asymptotic bulk region $X^0\to\infty$ is weakly coupled and weakly curved
(in fact
approximately supersymmetric), this means that back reaction is restricted to the intermediate region where the
tachyon $T$ is of order 1.

Third, the imaginary part of the 1-loop partition function is significant and will provide an important check on
the consistency of our computations. Recall that the analytic continuation \rotations\ included a rotation $\mu
= e^{-i\pi/2}\mu_E$.
This means that for real values of
our original parameter $\mu$,
the partition function has an imaginary part:
\eqn\Zdecomp{Z_1=\biggl(-{1\over\kappa} \ln  {\mu\over\mu_*}-i{\pi\over{2\kappa}}\biggr)\hat Z_1.}
Note that the IR cutoff $X^0_*$ is always real in our prescription.
We will interpret this as indicating that the system is in a pure state with thermal occupation numbers
corresponding to a temperature $\kappa/\pi$, a result which will also follow from an analysis of the 2 point
amplitudes at genus zero.  Let us return to this 
interpretation after explaining those computations in the next subsection.



In general, it would be interesting to unpack the 1-loop amplitudes
in more detail, to follow the fate of the
various closed string states and D-branes in our background.
An important aspect of this is mode mixing
induced by the tachyon operator:
the oscillator modes in the bulk generally mix
under the action of the tachyon
term in the region where it is substantial.

It might be possible to analyze
this using the ideas in \CohenSM.
In the type II case, a similar compuational
technique to the one we have described above
applies to the amplitudes of open strings in this
background, for example the 1-loop
open string partition function.
The closed string channel of such amplitudes
describes the response of the would-be graviton
and other closed string modes to D-brane sources. It is
necessary to specify consistently
the boundary conditions defining the D-branes
in this background, but it seems
likely that the $X^0_0$ integral
will again reveal that these amplitudes are shut off in the tachyon phase.
We note that
in the spacelike Liouville theory,
the ZZ-brane \ZZ\
is localized under the tachyon barrier,
and has a paucity of degrees of freedom.
It cannot move; basically it can only decay.
It would be very interesting to
understand the conformal boundary states in the
timelike case.

\subsec{2-point function, particle production, and Euclidean State}

Let us now include vertex operator insertions.  The $\mu$-dependence of amplitudes can be determined by a
similar technique to that above.  We analyze the derivative of the correlation function \EuclPI\foot{Note that
as in LFT, we use the semiclassical form of the vertex operators and dilaton as well as of the the action.} with
respect to $\mu_E$ by doing the integral over $X^0$'s zero mode $X^0_0$ first.
From that we can determine its dependence on $\mu_E$, and finally use
\rotations\ to determine its dependence on $\mu$.

This is similar to the above computation of the partition function, except now the integral over $y=e^{-\kappa
X^0_0}$ (which gave \yform\ in the case of the partition function) is of the form
\eqn\derG{{\del G(\{V_{n, -i\vec k_E}\})\over{\del\mu_E}}= \int [d\hat {\cal X}^0][d\vec{\cal X}][d({\rm ghost})]\int
dy  ~y^{- \sum_n i{\omega_n(\vec k_n)\over\kappa}}e^{-Cy} e^{-\hat S_E} {C \over \mu_E}  }
where $C$ is defined after \yform.
This yields a result for $G(\{V_{n, -i\vec k_E}\})$ proportional to
\eqn\npointmu{\mu_E^{i\sum_n \omega_n/\kappa}}
times a complicated path integral over nonzero modes, which would be difficult to evaluate directly,
but which is
independent of $\mu_E$. As discussed below \EuclacHet, we could rotate in a way that introduces a linear dilaton
in the Euclidean continuation. This would shift the $\omega$-dependent exponent by a term linear in $Q$, which
would rotate back to zero
when
we rotate $\mu_E$ back to $\mu$.

Fortunately, in the case of the 2-point function, we can use a simple aspect of the analytic continuation we used
to define the path integral to determine the magnitude of the result. As explained for example in \StromGut, the
two-point function of two negative frequency modes in the bulk is
\eqn\twointerp{G(\vec k, n;\vec k', n')=\delta_{nn'}\delta(\vec k-\vec k'){\beta_{\vec k, n}\over\alpha_{\vec k,
n}}}
where $\alpha_{\vec k, n}$ and $\beta_{\vec k, n}$ are the Bogoliubov coefficients describing the mixing of
positive and negative frequency modes.  This is the timelike Liouville analogue of the reflection coefficients
describing the mixing of positive and negative momentum modes bouncing off a spacelike Liouville wall.

In fact, this relation is precise here, and we can determine the magnitude $| \beta_\omega/\alpha_\omega|$ as
follows. As we discussed above for the partition function, after performing the Euclidean path integral defined
via the rotations \rotations, we must continue back $\mu \to - i\mu$ in order to obtain the amplitude
of interest. The regions where the worldsheet potential is positive translate in the Euclidean path
integral to a positive Liouville wall.  For these regions, the Euclidean 2-point function is a reflection
coefficient of magnitude 1.
The continuation above
\rotationsHet\rotations\ in $\mu$,
\eqn\againmu{\mu =  e^{-i{\pi\over 2}}\mu_E }
therefore yields a 2-point function for the Lorentzian theory of magnitude
\eqn\twomag{\Big |{\beta_{\vec k, n}\over\alpha_{\vec k, n}}\Big |= e^{-\omega(\vec k, n)\pi/\kappa} .}

Using the relations $|\alpha_\omega|^2 \mp |\beta_\omega|^2=1$
for bosonic and fermionic spacetime fields,
and the fact that the number of particles produced
$N_{\vec k, n}$ is given by $|\beta_{\vec k, n}|^2$,
this result translates into a distribution of pairs of
particles of a thermal form
\eqn\numberdist{N_{\vec k, n}={1\over{e^{2 \pi \omega(\vec k, n)/\kappa}\mp 1}} . }
This corresponds to a Boltzmann suppression of the distribution of pairs of particles (each pair having energy
$2\omega$) by a temperature ${ T}=\kappa/\pi$. This temperature can also be deduced from the imaginary part of
the 1-loop partition function \Zdecomp, providing a check on the calculations, as follows.

At the level of the genus zero two-point functions, the system is in a pure state whose phase information we
have not computed, but whose number density is thermal. We would like to understand 
how the result \Zdecomp\
fits into this thermal interpretation.  To explain this, for simplicity let us work in a field theory limit in
the bulk, so that we can express the state in terms of the bulk Fock space of states built on the bulk vacuum
$\ket{{\rm bulk}}$ killed by the annihilation operators $a_{\vec k}$. 
Ignoring interactions,
the bulk
state is of the form of a squeezed state
\eqn\thermalstate{\ket{\Psi_{0}}={\cal N} \exp{\left(
\sum_{\vec k} 
e^{i \gamma_{\vec k}}e^{-\omega(\vec k) \kappa/\pi}a^\dagger_{\vec
k}a^\dagger_{-\vec k}\right)} \ket{{\rm bulk}}}
where ${\cal N}$ is a normalization factor and the $\gamma_{\vec k}$ are the phases arising in the two point
function. Expanding the exponent, this is of the form
\eqn\thermalexpanded{\ket{\Psi_{0}}={\cal N} \sum_{n} e^{i\gamma_n} e^{-{\beta_T\over 2}E_n} \ket{E_n}}
for some real phases $\gamma_n$, where $n$ indexes the Fock space states and $E_n$ the corresponding energies,
and where $\beta_T=\pi/\kappa$ is the inverse temperature.

Switching back to the Schrodinger picture for simplicity, the bulk state can be thought of then as evolving as
if the time evolution operator, normally $U(t,0)=T\left(\exp{\left(-i\int_0^t dt' H(t')\right)}\right)$, 
is now
$T\left(\exp{\left(-i\int_{C}dt'H(t') \right)} \right)$ where the contour $C$ runs from 0 to $t$ along the real axis, and then runs
vertically from $t$ to $t-i\beta_T/2$.\foot{This is the first half of the contour obtained in real time thermal
field theory in a {\it mixed} thermal state; see for example section 5 of the second reference in \Steve\ for a
recent discussion.}

Now let us incorporate the 1-loop vacuum energy correction \Zdecomp.  This calculates the zero point energy
contribution to the spacetime Hamiltonian, integrated over time.  As we have just seen, in a state with thermal
occupation numbers, the effective Hamiltonian evolution up to time $t$ is on a time contour with an extra
segment from $t$ to $t-i\beta_T/2$.  The result \Zdecomp\ arises from the bulk vacuum result via exactly such a shift,
with $\beta_{ T}=\pi/\kappa$ corresponding to a temperature ${T}=\kappa/\pi$. Namely, as discussed in \S3.2, the
partition function is the summed zero-point energy in spacetime, times the volume of time: $Re(Z)=\Lambda
V_{d-1}V_0$. Said differently, $Z=\int dt \Lambda V_{d-1}$ where $t$ is the 
target-space time direction.
The
imaginary part of our amplitude \Zdecomp\ is obtained from the bulk vacuum by shifting
the spacetime Hamiltonian evolution by $\Lambda V_{d-1}X^0 \to \Lambda V_{d-1}(X^0-i{\pi\over{2\kappa}})$. So
the same temperature arises in both the genus zero and the 1-loop corrected amplitudes in our prescription,
providing a check.

Altogether, this yields the following simple result.  Let us consider the big bang case, with the tachyon
condensate turned on in the past.   Modulo expected subcritical tachyons in the type II case, the closed string
states are lifted in the far past (and in the type II case, we expect the subcritical tachyons to also condense
and lift degrees of freedom). Start with no excitations above this tachyon background (perhaps a natural choice
given the enormous effective masses in this region). The state in the bulk $X^0\to\infty$ region has a thermal
distribution of pairs of particles \numberdist, with temperature $\kappa/\pi$.  These pairs are created during
the phase where the tachyon condensate is order one\foot{ Indeed, the time-dependence of the Hamiltonian is only
non-adiabatic $1 \sim {\dot \omega \over \omega^2} ={\mu^2 \kappa e^{-\kappa t} \over \omega^3} $ in a small
window of time near $ t \sim 1/\kappa $. Similar suppression obtains for other measures of nonadiabaticity $
{\del_t^n \omega\over \omega^{n+1}} {\buildrel {t \to -\infty} \over \longrightarrow} (\kappa / \mu)^n e^{ - n
t/2}$ .}, and hence the calculation is self-consistent if we tune the bare dilaton to weak coupling.

This choice of state is analogous to the Hartle-Hawking, or Euclidean, State in the theory of quantum fields on
curved space, but it arises here in a perturbative string system via crucially stringy effects.  In quantum
field theory on curved space, the Euclidean vacuum is obtained by calculating Greens functions in the Euclidean
continuation of the spacetime background (when it exists) and continuing them back to Lorentzian signature. In
our case, a similar continuation has been made, but here the Euclidean system is a spacelike Liouville field
theory. The choice of vacuum (nothing excited above the tachyon condensate) is natural from the point of view of
the spacelike continuation, as it
involves excitations of modes with
only one sign of frequency in the tachyon phase;
these Wick rotate to exponentially dying modes
under the Liouville wall.

\subsec{Singularity Structure}

So far we focused on two particularly instructive physical quantities:  the 1-loop partition function and the
genus zero two-point function (Bogoliubov coefficients).  Let us now determine the singularity structure of more
general amplitudes. This is important in order to complete our assessment of the ability of the tachyon
condensate to resolve the spacelike singularity.  Namely, if the perturbative amplitudes are finite up to
expected divergences associated with physical states (which we will make precise below), then we may conclude
that the perturbative string theory is capable of resolving the singularity in the circumstances we have
specified (in particular, at weak coupling).

\subsubsec{N-point functions at genus zero}

As discussed above, the genus zero two-point function describes particle production in the linearized spacetime
theory.  The singularity structure of general $N$-point amplitudes can be ascertained from the path integral
\EuclPIHet\EuclPI.  In a nontrivial bulk vacuum, such as that derived above in the Euclidean vacuum \numberdist,
we are interested in a linear combination of vertex operators \superVHet\
comprised of $\alpha_{\vec k, n}$ times
a negative frequency component plus $\beta_{\vec k, n}$ times a positive frequency component.

The path integral diverges when a bosonic degree of freedom can go off to infinity unobstructed by the
$e^{-S_E}$ factor.  As discussed in the introduction, this situation appears in the big bang region of the
spacetime in the naive extrapolation of GR, as the space shrinks and the kinetic terms in $S_E$ go away.  In our
case, where it is positive the tachyon term obstructs this divergence (everywhere in the Heterotic case, and
away from the subcritical type 0 regions of the type II system).

There are divergences in the bulk region $X^0\to\infty$ that are expected in a time dependent S matrix.
Generically, the vertex operators provide oscillations suppressing the path integral contribution in this
region. However a divergence in $X^0_0\to +\infty$ appears when the frequencies are such as to cancel this
oscillation:
\eqn\singstructure{\sum\omega_n(\pm 1)_n = 0}
The $\pm$ sign here comes from the presence of both positive and negative frequency modes in the vertex
operators.\foot{This is another aspect analogous to the situation in Liouville theory
(see eqn.\ (87) of
\teschner).} These divergences correspond to expected divergences from physical intermediate states in time
dependent systems (see \eg\ \BD\ chapter 9 for a discussion of this).

In \refs{\StromTak,\StromGut,\SchomerusVV}\ subtleties were encountered in a prescription for analytically
continuing higher point amplitudes from spacelike Liouville theory.  It will be interesting to study these
issues in more detail in our case, keeping careful track of the vacuum in which the computations are being done
in the various prescriptions for continuing the amplitudes.

\subsubsec{Higher loops}

Higher loop amplitudes arise from the path integral \EuclPIHet\EuclPI\ defined on a Riemann surface of higher
genus $h$. These contain dependence on the dilaton $\Phi\equiv \Phi_0+ \hat\Phi$  where $\Phi_0$ is the constant
value in the bulk region.  This introduces a factor of $e^{\Phi_0(2h-2)}$ from the bare bulk string coupling as
well as a contribution
\eqn\dilterm{S_\Phi = \int_\Sigma \hat{\cal R}^{(2)} \hat\Phi[X^0] }
(plus its supersymmetric completion in the Heterotic and Type II cases).  Semiclassically $\Phi =\Phi_0$ (\ie\
$\hat \Phi=0$) as discussed in \dilsemi.   The dilaton will ultimately
get sourced by the tachyon. The
corresponding corrections will be generated by the worldsheet path integral, but are suppressed by powers of
$e^{\Phi_0}$.  Moreover, as in our analysis of the 1-loop vacuum amplitude, the $X^0_0$ integral reveals that
higher genus amplitudes have support limited to the weakly coupled bulk of spacetime.

\newsec{Positively-curved spatial slices}

In this section,
we generalize our techniques to strings in geometries of the form \GRsing\
where the $\Omega$ are coordinates on higher dimensional spheres.
The worldsheet theory
will be described by an $O(N)$ model
at an energy scale related to $X^0$
in a way we will specify.
In this case there is no
topologically-stabilized winding tachyon\foot{
It might be interesting to consider examples
of positively-curved spaces with
nonzero $\pi_1$ such as $S^N/\Gamma$ with freely acting $\Gamma$.
}.
The
sigma model on spatial slices
nevertheless develops a mass gap.
We will frame this fact
in terms of the discussion of \S1,
and investigate the extent to which it
can be used to remove the
singularity present in the GR approximation.
Some aspects of the analysis of \S3\ persist.
Unlike in the case of flat spatial slices, however,
the back-reaction from the velocity of the radion
will be harder to control in these examples.

\subsec{The mass gap of the $O(N)$ model}

%

%

Consider $N$ two-dimensional
scalar fields arranged into an $O(N)$ vector $\vec n$.  The partition function
of the $O(N)$ model is
\eqn\npartitionone{ Z = \int [dn] e^{ - \int d^2z {R^2} ( \del_\mu \vec n)^2}
\prod_{z} \delta( n^2(z) - 1) }
A nice way to see the mass term
appear is to use a Lagrange multiplier to enforce the delta function localizing
the path integral onto a sphere, and large $N$
to simplify the resulting dynamics (see \eg\ \NovikovAC):
\eqn\npartitiontwo{ Z = \int [dn] \int [d\lambda]
e^{ - \int d^2z [{R^2}  \vec n ( -\del^2 + i \lambda ) \vec n  + i \lambda ] }
}
where $\lambda $ is the Lagrange
multiplier field introduced to represent the delta function.
Now integrate out $n$:
\eqn\npartitionthree{ Z = \int [d\lambda] e^{ - N/2 \tr \ln ( - \del^2 + \lambda )
+ {R^2} \int d^2 z \lambda } .}
At large $N$, the $\lambda$ integral has a
well-peaked saddle at
\eqn\lambdavev{ \lambda(x) = - i m^2 }
where the mass $m$ satisfies
\eqn\saddleeqn{ { R^2} = N \int^\Lambda {d^2k \over (2\pi)^2} {1 \over k^2 + m^2}
= { N \over 2\pi } \ln {\Lambda\over  m} .}
Renormalize
by defining the
running coupling at the scale $M$ by
\eqn\runningcoupling{ {R^2(M) } = {R_0^2} + {N \over 2\pi} \ln \Lambda / M.}
Plugging back into the action for $n$, we have a mass for the $n$-field which runs like
\eqn\effectivemass{ m = M e^{ - { 2\pi R^2\over N } } .}

An alternative UV completion of the model
which is sometimes more convenient
(and easier to supersymmetrize)
gives $\lambda$ a bare mass: add to the action
$$ \delta S = \int a \lambda^2 .$$
for a large parameter $a$.
Integrating out $\lambda$, this smoothens the delta function,
and imposes the $n^2 = R^2$ relation
weakly in the UV by a quartic potential.

\subsubsec{Supersymmetric $O(N)$ model}

Since we wish to study string theories
without bulk tachyons, we will need to
understand the supersymmetric version of the model.
A (1,1) supersymmetric version of the $O(N)$ model
has an action
$$ S = \int d^2 \theta  \left(
\epsilon_{\alpha\beta} D_\alpha n D_\beta n +   \Lambda ( n^2 - R^2 ) + a \Lambda^2 \right) ;$$
$ \alpha, \beta = \pm$,
$\Lambda = \lambda + \theta^\alpha \psi_\alpha +
\theta^2 F_\lambda $ is now a Lagrange multiplier superfield,
and
$ D_\alpha = {\del \over \del \theta^\alpha} + i \theta^\beta \sigma_{\beta \alpha}^\mu \del_\mu .$
Note that the type II GSO symmetry acts on $\Lambda$ as
\eqn\GSOonlambda{ (-1)^{F_L} : \Lambda \mapsto - \Lambda, ~~
(-1)^{F_R} : \Lambda \mapsto - \Lambda. }

The large $N$ physics is the same as in the bosonic
case (see \eg\ \MosheXN), exhibiting a mass gap,
except now there are two vacua for $\Lambda$.
When
\eqn\superlambdavev{ \vev{\lambda} = \pm m }
the GSO symmetry is spontaneously broken;
the two vacua are identified by the GSO projection.
This is just as in the appendix of \TFA,
and it results in
a single type zero vacuum.  This statement about the GSO
projection applies to all $N$, including the case $N=2$ of a shrinking
circle described in \S2, \S3.   In particular, this justifies the comment made
in the discussion above \Zderiv\ that the regions of negative potential in the type II worldsheet
have a type 0 subcritical GSO projection.


\subsec{$\IC\IP^1$ Model}

The $(1,1)$ sigma model on $S^2$ actually has $(2,2)$ supersymmetry.
Consider a $(2,2)$ linear sigma model \WittenYC\ for it.
There are two chiral superfields $Z^i$ each with charge one
with respect to a single $U(1)$ vectormultiplet.
The D-term equation is
\eqn\Dterm{ 0 = \sum_{i=1}^2 |Z^i|^2 - \rho . }
Below the scale $e$ of the gauge coupling,
this model describes strings propagating
on a 2-sphere of radius
$ R = \sqrt {\rho \alpha'} $.
The FI coupling $\rho$ flows logarithmically towards smaller values in the IR:
\eqn\FIRG{\rho(M) = \rho_0 - 2 \ln {M \over M_0} .}
This breaking of scale invariance is in the same $(2,2)$
supermultiplet as an anomaly in the chiral $U(1)$ R-symmetry;
only a $\IZ_2$ subgroup of this latter group is a symmetry of
the quantum theory (this is part of the
GSO symmetry in type II theories).

Integrating out the chiral multiplets $Z^i$
leads \WittenYC\
to an effective twisted superpotential for the vectormultiplet scalar
\eqn\effectivetwistedW{
\tilde W = 2 \Sigma \ln \Sigma - t \Sigma.}
Mirror symmetry \HoriKT\ relates this to a model with
one twisted chiral superfield $Y$, governed by a twisted superpotential
\eqn\twistedW{ \tilde W = \Lambda \left( e^Y + e^{-Y} \right) }
where $ \Lambda= m e^{ -t/2} , t = \rho + i \vartheta$.
This effective twisted superpotential
has isolated massive vacua.

Next let us discuss the GSO projection, to ensure that the relevant operator we
are generating is present in the type II theory (as opposed to being
a type 0 bulk tachyon).  In the type II case,
the twisted chiral superpotential must be odd under the chiral GSO actions.
This is accomplished by
\eqn\GSOonY{ (-1)^{F_L} : Y_2 \mapsto Y_2 + \pi , ~~ (-1)^{F_R}: Y_2 \mapsto Y_2 + \pi,}
where $ Y \equiv Y_1 + i Y_2 $,
or
simply $\Sigma\to-\Sigma$.

The twisted superpotential
\effectivetwistedW\
has two massive vacua
\eqn\sigmavacua{ \sigma = \pm e^{ -t/2} ,}
which are permuted by the GSO action.
The condensate \twistedW\
is therefore not a bulk tachyon mode.

We can use this mirror description to
further elucidate the physical interpretation of the
condensate.
It is invariant under the $SU(2) \simeq SO(3) $
rotations of the $S^2$.
Since $Y_2$ is the variable T-dual
to the phase of the $Z$s,
from the point of view of the
original linear model, \twistedW\ represents
a condensation of winding modes.
It is tempting to interpret this as a condensate resembling a ball of rubber-bands
wrapping great circles of the small sphere.

\subsubsec{A special RG trajectory}

In the case $N=3$ of the two-sphere,
where there is a two-cycle in the geometry,
there are topologically-charged worldsheet instantons.
The contribution
to the sum over maps
of the sector with winding number $n$
is weighted by $e^{i n \theta}$
where $ \theta = \int_{S^2} B $
is the period of the NSNS B-field through the two-sphere.

When $\theta=\pi$ this introduces wildly fluctuating
signs in the path integral
which can result
\FateevBV\
in a critical theory in the IR.
In fact, the model flows to
the $SU(2)$ WZW model at level one,
also known as a free boson at the self-dual radius.
{\it This} model
has a topologically-stabilized winding mode,
which is exactly massless.
At this point,
the evolution may be
glued onto the analysis
of \S2,3.

\subsec{Coupling to string theory}

{\it \hfill\hbox{Eternal nothingness is fine if you happen to be dressed for it. }

\hfill\hbox{-- Woody Allen}}

We need to check whether the mass gap
whose origin we have reviewed
takes effect {\it before} large curvature
develops.
In the example of \S2,3,
the rate of shrinking $ \del_t R $
of the circle was a tunable parameter
which we used to control the collapse.
In this case, where the spatial curvature exerts a force
on $R$, we will need to reevaluate the behavior of
$R(t)$.

In order to do this, we begin at large radius,
and use the fact that in this regime,
the beta function equations for the
worldsheet theory are the same as the
gravity equations of motion.
%

In the case of positive spatial curvature, the
Friedmann equation (the Hamiltonian constraint) requires a stress-energy
source which dominates over the
curvature contribution:
\eqn\Friedmann{\biggl({\dot R\over R}\biggr)^2=-{1\over R^2}+ G_N \rho}
%
where $\rho$ is the energy density in non-geometrical sources. The curvature term $-1/R^2$ alone, in the absence
of the term from extra sources $\rho$, would not yield consistent initial data; instead the source term must
dominate over the curvature term in the large radius general relativistic regime.  This means that unlike the
previous two cases of \S2.1,2.2, we do not classically have a tunable parameter allowing us to slow down the
approach to the would-be singularity.

\subsubsec{Inclusion of matter}

As we mentioned in \S2.3,
in the case of positive spatial curvature,
the Friedmann equation \Friedmann\
has no real solutions in the absence of matter.
We will overcome this problem
by including some nonzero radiation energy density on the RHS of \Friedmann.
With $ \rho = x/R^{N}$
($x$ is a constant and
$N$ is the number of spacetime dimensions participating in the FRW space),
the maximum radius reached is
\eqn\maxradius{ R_{{\rm max}} = l_P x^{1\over N-2} ,}
where $l_P = G_N^{1\over N-2}$ is the $N$-dimensional Planck length.
In the curvature-dominated regime,
\eqn\curvaturedominated{ R(t) \sim R_{{\rm max}}^{N-2 \over N} t^{2 \over N} .}

Now we can estimate the time at which the mass gap takes effect. For convenience (as opposed to phenomenology),
consider the case $N=4$, where \curvaturedominated\ implies $R(t) \sim \sqrt {tR_{max}}$. Semiclassically, the
tachyon term in the worldsheet effective action \effectivemass\ depends on time via the ``tachyon" profile
\eqn\tachyonprofile{ \CT(t) \sim \mu e^{ - R^2/N \alpha' } . } If we assume that the leading effect of the
radiation that we added
is to fix the evolution of the scale factor (\ie\ that it does not couple significantly to
the Liouville mode in any other way), we can make a similar estimate to those of the previous sections. In fact,
for $N=4$, the zeromode integral over $X^0$ is of the same form as \Zresult \eqn\truncationagain{ Z \propto -
\ln \mu .}

$X^0$ goes to zero at the would-be bang singularity.  The
range of $X^0$ for which the amplitudes have support is $X^0 > {1\over T}
 \ln \mu/\mu^*$.  Increasing $\mu$ makes the
range of $X^0$ support of amplitudes smaller.
So if we take $\mu$ to be large, we can ensure that the lifting
of modes occurs in a regime where the kinetic terms have not yet died as we approach the singularity.

Physically, this parameter $\mu$ determines the amplitude of the oscillating mode in the bulk and hence its
initial behavior in its exponential regime.  We are introducing a classical solution with a large amplitude
condensate of tachyon even in the initial ``bulk" region where the space is larger than string scale, but we
expect that these modes will decay once the other states come down.

\subsubsec{Particle Production}

From here the analysis proceeds as in \S3, but the absence of
a tunably small rate of growth of the $S^3$ space leads to
a much larger density of produced closed strings.
In particular,
we again obtain an effective temperature
via the periodicity in imaginary time of the condensate.
Using the definition
$$ \CT(t) \propto e^{ - T t} $$
for the effective temperature $T$,
we find (again, for $N=4$)
$$  T \sim {R_{max}  \over \alpha' } .$$
Thus, when the cosmology has a phase
during which it is bigger than string scale,
the effective temperature
is larger than the Hagedorn temperature.
This is in contrast to the tunably small value of $\kappa$
we obtained in \numberdist\ in the case of flat spatial slices.

The upshot is that in this case of positively curved spatial slices, although the mass
gap lifts the would-be GR divergences in the worldsheet path integral, a new
source of back reaction is generated through copious particle production.
It is worth emphasizing that the GR solution alone will lead to particle
production of momentum modes, whose back reaction may also correct
the background in an important way.  We leave this analysis and its potential
application to Schwarzchild black hole physics to further work \usnext.

\newsec{Discussion}

\subsubsec{Application to Black Hole Physics}

Spacelike singularities appear inside generic black hole solutions of general relativity.  The case of a
shrinking $S^2$ described in \S4.1\ appears inside the horizon of the Schwarzchild black hole solution in four
dimensions (with an additional spatial direction $t$ which is stretching at the same time)
\eqn\Schwarzchild{ds^2=-(1-L_S/r)dt^2+{dr^2\over{1-L_S/r}}+r^2d\Omega^2}
where $L_S$ is the Schwarzchild radius.  Inside the horizon ($r<L_S$), $r$
is a timelike coordinate.

When the $S^2$ parameterized by $\Omega$ shrinks, the worldsheet path integral develops contributions arising
from the mass gap of the corresponding sigma model as discussed in \S4.1. It would be interesting to understand
if this might clarify black hole dynamics \usnext.   These results may also apply to the proposal of
\BHfinalstate, where the possibility of postselection on a ``nothing" state was explored.  The unitarity
required in \BHfinalstate\ may arise from the unitary evolution along the $t$ direction inside the horizon,
generated by the momentum generator inside the horizon.

\subsubsec{Other vacua and the shape of the S-matrix}

We have focused on a vacuum with no extra excitations above the tachyon background in the initial state.  This
is motivated by the lifting of closed string degrees of freedom in the presence of the tachyon.  However, it
would be very interesting to understand if other states are allowed.  The analogy with quantum field theory with
exponentially growing mass suggests a similar ``massing up" of closed string modes in tachyon backgrounds in
more general states.  Moreover some interesting potential consistency conditions which may restrict the range of
allowed states suggest themselves.  In particular, a propagator entering the tachyon phase (rather than
annihilating against another mode) has the property that semiclassically its worldsheet time ends at a finite
value as $X^0$ goes to infinity.  A similar issue arises in formulating the S matrix in the case of pure quantum
field theory with an exponentially growing mass (see \Schomerusanomaly\ for a discussion of the issues there);
in a generic vacuum there is a sense in which time ends prematurely in this system due to the absence of well
defined on shell asymptotic states. Some aspects of the specification of the spacetime quantum state from the
worldsheet point of view have been discussed in \refs{\Strompart, \branedecay}.

In particular, one of the main questions raised by spacelike singularities is that of predictivity.  In field
theory or GR on a background with a putative big bang singularity, the initial conditions on the fields in the
bulk region are ambiguous.  If there are other consistent states of the system involving some extra excitations
introduced initially and becoming light as the tachyon turns off, then the singularity, while resolved, will not
be arbitrarily predictive.  It is important to understand the status of all possible states.

\subsubsec{Big crunch}

Our main computations were done in the vacuum discussed in \S3\ with
no excitations above the tachyon condensate.   In the case of the big
bang, this is perhaps a natural choice of initial state.
In the case of the big crunch,
the methods employed in this paper
are not yet sufficient to
answer the question
of what happens starting from an arbitrary initial state.
For example, it
is interesting to ask what happens if we start with no particles in the bulk.
At the level of the genus zero diagrams, we can accomplish this by considering
correlation functions of vertex operators which are nontrivial linear combinations
of positive and negative frequency modes in the bulk.  For the 1-loop and higher
genus diagrams, it is an open question here (as in the case of open string
tachyons) how the different vacua translate into different prescriptions for
the worldsheet path integral.

One aspect of the system is pair production of winding
modes themselves as they become massless \BerkoozRE;
this can drain energy from the rolling radius to some
extent \trapping.

\subsubsec{Negative spatial curvature}

We have discussed the cases of $k=1$ and $k=0$.
It is natural to ask about the case  $k=-1$ where
the spatial sections have negative curvature.
In this case, at large radius $R$, the system expands according to
the simple relation $R\sim X^0$.   Localized tachyon dynamics
in some such examples were discussed in \TFA.

\lref\Saltman{
  A.~Saltman and E.~Silverstein,
  ``A new handle on de Sitter compactifications,''
  arXiv:hep-th/0411271.
}

The big bang singularity in the far past in this case
is not related by RG flow toward the IR in the matter sector
of the corresponding worldsheet sigma model.  The direction
of flow is opposite;  the small radius big bang
regime corresponds to the UV.   Hence in this case, the big
bang resolution may depend on the appropriate UV completion
of the sigma model on negatively curved spatial slices.\foot{One
can alternatively add ingredients to metastabilize the system away from this
difficult regime \Saltman.}

\subsubsec{Cosmology}

It will be interesting to see if these methods and results translate into concrete results for more realistic
string cosmology.  Inflation tends to dilute information about the big bang singularity, but depending on the
level of predictivity of the singularity, it may nonetheless play a role.  Stretched strings play an important
role in our mechanism for resolving the singularity:  perhaps there is some relation between them and late-time
cosmic strings (whether inside or outside the horizon at late times).

\subsubsec{Toward a theory of Nothing}
{\obeylines
{\it \hfill\hbox{It is the silence between the notes that makes the music;}
\hfill\hbox{it is the space between the bars that holds the tiger.}
\hfill\hbox{-- Anonymous}}
}

Our calculations using methods borrowed from Liouville theory
exhibit truncation of support of amplitudes to the bulk of
spacetime, and hence concretely support the notion of a ``Nothing"
phase in the regime of the tachyon condensate.  Conversely,
spacetime emerges as the tachyon turns off.

It would be
very interesting to characterize this phase and its onset
in more detail,
for example by unpacking the partition function to analyze
its individual contributions.  Although perturbative
methods exhibit the basic effect,
perhaps there is some dual
formulation for which the emergence of time
as the tachyon turns off is also built in.

\bigskip
\bigskip
\centerline{\bf{Acknowledgements}}

We would like to thank A. Adams, M. Berkooz, S. Kachru, X.~Liu, A. Maloney, H. Ooguri, J.~Polchinski, N.
Seiberg, S. Shenker, A. Strominger for helpful discussions, and Gary Horowitz for early collaboration and very
helpful discussions. We are supported in part by the DOE under contract DE-AC03-76SF00515 and by the NSF under
contract 9870115.

\listrefs

\end

We can interpret this as indicating that the system is in a thermal state, as follows.  A thermal system is
described in a real-time formalism by shifting time by $i$ times half the inverse temperature:  $t\to
t-i\beta_{T}/2$.
The result \ZdecompII\ arises from the bulk vacuum result via such a shift, with $\beta_{ T}=\pi/\kappa$
corresponding to a temperature ${T}=\kappa/\pi$.  Namely, as discussed above, the partition function is the
summed zero-point energy in spacetime, times the volume of time: $Re(Z)=\Lambda V_{d-1}V_0$. Said differently,
$Z=\int dt \Lambda V_{d-1}$ where $t$ is the time direction in spacetime. The imaginary part of our amplitude
\Zdecomp\ is obtained from the bulk vacuum by shifting
the spacetime Hamiltonian evolution by $\Lambda V_{d-1}X^0 \to \Lambda V_{d-1}(X^0-i{\pi\over{2\kappa}})$.

In the next section, we will perform a check of this result by showing that our path integral defines the theory
in a
state with thermal occupation numbers
in the bulk.  In particular, we will calculate the magnitudes of the tree-level two-point amplitudes (as well as
the $\mu$-dependence and singularity structure of higher point amplitudes).  We will determine from these
amplitudes the magnitude of Bogoliubov coefficients describing particle production in the bulk; the result will
be that if we start with no excitations in the far past, we obtain a thermal distribution of pairs of closed
strings with this temperature $\kappa/\pi$.

A thermal system is described in a real-time formalism by shifting time by $i$ times half the inverse
temperature:  $t\to t-i\beta_{T}/2$.
The result \Zdecomp\ arises from the bulk vacuum result via such a shift, with $\beta_{ T}=\pi/\kappa$
corresponding to a temperature ${T}=\kappa/\pi$.  Namely, as discussed above, the partition function is the
summed zero-point energy in spacetime, times the volume of time: $Re(Z)=\Lambda V_{d-1}V_0$. Said differently,
$Z=\int dt \Lambda V_{d-1}$ where $t$ is the time direction in spacetime. The imaginary part of our amplitude
\Zdecomp\ is obtained from the bulk vacuum by shifting the zero point energy part of the spacetime Hamiltonian
evolution by $\Lambda V_{d-1}X^0 \to \Lambda V_{d-1}(X^0-i{\pi\over{2\kappa}})$.

In the next section, we will perform a check of this result by showing that our path integral defines the theory
in a
state with thermal occupation numbers
in the bulk.  In particular, we will calculate the magnitudes of the tree-level two-point amplitudes (as well as
the $\mu$-dependence and singularity structure of higher point amplitudes).  We will determine from these
amplitudes the magnitude of Bogoliubov coefficients describing particle production in the bulk; the result will
be that if we start with no excitations in the far past, we obtain a thermal distribution of pairs of closed
strings with temperature $\kappa/\pi$.
